\documentclass[prd,twocolumn,superscriptaddress,floatfix,showpacs,nofootinbib]{revtex4-2}
\usepackage{amsfonts}
\usepackage{amsmath}
\usepackage{amssymb}
\usepackage{xcolor}
\usepackage{graphicx}
\font\mybb=msbm10 at 10pt
\def\bb#1{\hbox{\mybb#1}}
\def\be{\begin{equation}}
\def\ee{\end{equation}}
\def\bea{\begin{eqnarray}}
\def\eea{\end{eqnarray}}

\begin{document}
\begin{flushright}
30/12/2022, 
V3: 
10/04/2023
\end{flushright}
\bigskip
\title{
Properties of multiple D$0$-brane system: 11D origin, equations of motion and their solutions
}

\author{Igor Bandos}
\email{igor.bandos@ehu.eus}
\affiliation{Department of Physics and EHU Quantum Center, University of the Basque Country UPV/EHU,
P.O. Box 644, 48080 Bilbao, Spain,}
\affiliation{IKERBASQUE, Basque Foundation for Science,
48011, Bilbao, Spain, }

\author{Unai D.M. Sarraga$^1$}
\email{unai.demiguel@ehu.eus}

\bigskip

\begin{abstract}
We study the properties of 10D  multiple D$0$-brane (mD$0$) system described by recently proposed complete supersymmetric and $\kappa$-symmetric nonlinear action which includes an arbitrary positive definite function ${\cal M}({\cal H})$ of the relative motion Hamiltonian ${\cal H}$.
First we show how the action with a particular nonlinear  ${\cal M}({\cal H})$ can be obtained from the action for 11D multiple M-wave (multiple M$0$-brane or mM$0$) system. Then we obtain the complete set of equations of motion of mD$0$ system with arbitrary positive definite ${\cal M}({\cal H})$, perform a convenient gauge fixing, solve the center of energy equations and establish an interesting correspondence between  the relative motion mD$0$ equations and the equations of maximally  supersymmetric SU($N$) Yang-Mills theory (SYM). We show that this correspondence does not imply a gauge equivalence but establishes a relation between solutions of the system. In particular,  it implies that all the supersymmetric solutions of mD$0$ equations in its relative motion part coincide with supersymmetric solutions of the SYM equations.

\end{abstract}

\maketitle

\section{Introduction}

Supersymmetric extended objects, super-$p$-branes\footnote{Here $p$ refers to the number of spacial dimensions of the worldvolume of the object so that $p=1$ corresponds to strings, $p=2$ corresponds to membrane and 0-branes are supersymmetric particles. }, play very important role in String/M-theory \cite{Polchinski:1998rq,Polchinski:1998rr,Becker:2006dvp,West:2012vka}  and in AdS/CFT duality
which has been developed  to a much more general gauge-gravity correspondence. Particularly interesting are
ten dimensional ($\text{D}=10$) Dirichlet $p$-branes or D$p$-branes the worldvolumes of which are the surfaces where the fundamental string (sometimes called F$1$-brane) can have its ends.

The (theoretical) discovery of these objects is dated by late 80$^{{\rm th}}$ \cite{Sagnotti:1987tw,Dai:1989ua,Horava:1989ga} but the common interest to them was awakened  by the famous paper by Polchinski \cite{Polchinski:1995mt} in which he identified them with that time looked mysterious supersymmetric solutions of ${\cal N}=2$ (type IIA and type IIB) supergravity and  showed that they carry the charges of Ramond-Ramond q-form gauge fields of these supergravity theories. The complete nonlinear worldvolume actions for the D$p$-brane possessing worldline supersymmetry and
the local fermionic $\kappa$-symmetry was found soon by several groups
\cite{Cederwall:1996pv,Aganagic:1996pe,Cederwall:1996ri,Aganagic:1996nn,Bergshoeff:1996tu,Bandos:1997rq}
but even before the supersymmetric equations of motion had been found by Howe and Sezgin \cite{Howe:1996mx} in the frame of superembedding approach \cite{Bandos:1995zw,Sorokin:1999jx}.

The spacetime filling D9-brane in type IIB supergravity superspace corresponds to the case of open superstring the ends of which can move freely in the (super)spacetime (see \cite{Green:1987sp} and refs therein). Such an end of the string can be coupled to a vector gauge field. The vector gauge field is also present in the quantum state spectrum of the open string, so that in this case it is natural to assume that we are dealing with coupling of string to one of its excitation considered as a backgrounds. The equation of motion for such a background gauge field can be found from the requirement of preservation of the conformal symmetry of string theory in such a background (vanishing of the beta functions). Those was found \cite{Fradkin:1985qd} to be the same as followed from the Born-Infeld action  \cite{Born:1934gh} which thus can be considered as  (the bosonic limit of) an  effective action for the open (super)string with free ends and, in modern language, the effective action for the spacetime filling D9-brane\footnote{Also in the case of bosonic string considered in \cite{Fradkin:1985qd} one speaks about spacetime filling D25-brane, we allow ourselves to streamline the discussion a bit referring directly to the generalizations for the case of superstring. }. Indeed, it was shown that D$p$-branes with $p\leq 9$ carry out worldvolume vector gauge fields and thus that the effective action for this is the $\text{d}=p+1$ dimensional Dirac-Born-Infeld action (which is to say the generalization of Born-Infeld action to curved space with metric of this induced by the embedding of the worldvolume in target spacetime) \cite{Leigh:1989jq}.

Interestingly enough, from the early years of string theory it was known that the ends of open string can be supplied with additional degrees of freedom describing ``isospin'' \cite{Paton:1969je}. This fact was  associated to  the possibility to attach ``quarks'' to the ends of string.
In the frame of Lagrangian mechanics of the string model those are described by auxiliary fermionic fields (boundary fermions) ``living'' at the string endpoints \cite{Bars:1975dd,Marcus:1986cm}. In the case of string amplitudes the ``isospin'' degrees of freedom manifest themselves  as additional factor constructed from the trace of matrices representing the generators of a non-Abelian internal symmetry group which in the simplest case is considered to be U$(N)$ \cite{Paton:1969je}.

From this point of view one can find natural the brilliant idea of Witten  \cite{Witten:1995im} that in the limit  of $N$ (nearly) coincident D$p$-brane, when the strings connecting different  branes can be described by (nearly) massless U(1) gauge fields, as the string with both ends on the same brane can, the manifest U$((1))^{N^2}$ symmetry of the system is enhanced till U$(N)$ and thus in low energy limit (and after an appropriate gauge fixed with breaking Lorentz invariance) shall be described by non-Abelian supersymmetric Yang-Mills (SYM) model.

The search for a (more) complete nonlinear multiple D$p$-brane (mD$p$) action describing the system of $N$ nearly coincident D$p$-branes and strings ending on different and the same branes has more than 25 years of history. Despite many interesting and deep results obtained on this way
\cite{Tseytlin:1997csa,Emparan:1997rt,Taylor:1999gq,Taylor:1999pr,Myers:1999ps,Bergshoeff:2000ik,
Bergshoeff:2001dc,Sorokin:2001av,Janssen:2002vb,Drummond:2002kg,Panda:2003dj,Janssen:2002cf,Janssen:2003ri,
Lozano:2005kf,Howe:2005jz,Howe:2006rv,Howe:2007eb,Bandos:2009yp,Bandos:2009gk,Bandos:2010hc,McGuirk:2012sb,
Bandos:2012jz,Bandos:2013uoa,Choi:2017kxf,Choi:2018fqw,Bandos:2018ntt,Brennan:2019azg,Bandos:2021vrq,Bandos:2022uoz}
one cannot say that the problem is definitely solved even in the simplest case of mD$0$-system.
The expected properties of this action, which should be a kind of non-Abelian generalization of the single D$p$-brane actions from \cite{Cederwall:1996pv,Aganagic:1996pe,Cederwall:1996ri,Aganagic:1996nn,Bergshoeff:1996tu,Bandos:1997rq}, are  Lorentz invariance, spacetime (target superspace) supersymmetry, the counterpart of local fermionic $\kappa$-symmetry of single D$p$-brane (which makes the ground state of the system supersymmetric) and appearance of the U($N$) SYM action as low energy description in the suitable gauge.

The so-called ``dielectric brane'' action by Myers \cite{Myers:1999ps}, which was widely accepted as a bosonic limit of such an action and was studied and generalized in \cite{Janssen:2002vb,Janssen:2002cf,Janssen:2003ri,Lozano:2005kf}, still resists its supersymmetric generalization\footnote{The bilinear fermionic contributions to the hypothetical supersymmetric generalization of the action \cite{Myers:1999ps} were searched for in \cite{McGuirk:2012sb} using the suggestion form the studies of single D$p$-brane action in \cite{Martucci:2005rb}.
Notice that the approach with non-Abelian version of $\kappa$-symmetry \cite{Bergshoeff:2000ik} successful in the linearized approximation cannot be extended to cubic level as it was shown in  \cite{Bergshoeff:2001dc}.} (neither Lorentz invariant formulation of this is known). An interesting approach on ``-1 quantization'' level was developed in \cite{Howe:2005jz,Howe:2006rv,Howe:2007eb}. It uses the superembedding approach version of   boundary fermion formalism and it looks like to obtain the non-Abelian counterpart of the complete nonlinear action for single D$p$-brane on this basis the problem of quantization of the system including supergravity and ``-1 quantized'' D$p$-brane should be addressed (more discussion on this approach can be found in \cite{Bandos:2018ntt}).

The superembedding approach to mD$0$ and related mM$0$ (multiple M-wave or multiple M$0$-brane) system in type IIA 10D and in 11D superspaces was developed in \cite{Bandos:2009yp,Bandos:2009gk,Bandos:2010hc}. It was based on the standard superembedding equations, the same as was used to describe single D$0$-brane,  and seemed to provide some approximation to the mD$0$ and mM$0$ equations. The suggestion on how to possibly modify the superembedding equations might come from the action principle and the action for 11D mM$0$ system was constructed in \cite{Bandos:2012jz} and studied in \cite{Bandos:2013uoa}.

In \cite{Bandos:2022uoz} we have constructed the nonlinear doubly supersymmetric action which possesses the properties expected for 10D mD0 brane system
 (its $D=3$ counterpart had been found in \cite{Bandos:2021vrq}). An unexpected property was the presence in the action of an arbitrary positively definite function ${\cal M}({\cal H})$ of the relative motion Hamiltonian ${\cal H}$ (constructed from the matrix fields of the $\text{d}=1$ ${\cal N}=16$ SU($N$) SYM supermultiplet)\footnote{Similar property is possessed by the 0-brane model proposed in \cite{Panda:2003dj}. See  \cite{Bandos:2018ntt} for the discussion on this model in the perspective of our approach.}. The model corresponding to constant ${\cal M}({\cal H})=m$ has been constructed before in \cite{Bandos:2018ntt} by adding to the single D$0$-brane action (describing then the center of mass motion of mD0 system) of the action for  1d ${\cal N}=16$ SU($N$) SYM supermultiplet coupled to supergravity induced by the embedding of the center of mass worldline into the target type IIA superspace. The problem of the relation of this candidate mD0 action to the previously known action for 11D multiple M0-brane (multiple M-wave or mM0) system from \cite{Bandos:2012jz} was posed but was not solved in  \cite{Bandos:2018ntt}.

In this paper we will show how the dimensional reduction of the mM$0$ action, performed after a suitable redefinition of the matrix fields, produces a specific representative of the family of candidate mD$0$ action from \cite{Bandos:2022uoz} with a particular form of the function
${\cal M}({\cal H})$:
\be\label{cM=m+}
{\cal M} = \frac m 2 +  \sqrt{\frac {m^2} {4}+\frac {\cal H} {\mu^6}}\; ,
\ee
where $m$ and $\mu$ are the constants of dimension of mass. The first of these is associated with the center of mass motion while the second is the coupling constant of the relative motion and center of energy variables.

Then we study the equations of motion which follows from the action of \cite{Bandos:2022uoz}. In particular we show that equations of motion imply  the preservation of the relative motion Hamiltonian, d${\cal H}=0$ and, hence, of the on-shell value of the mass of the mD$0$ system
\be
\mathfrak{M} = m + \frac{2}{\mu^6} \frac{\mathcal{H}}{\mathcal{M}(\mathcal{H})}\,\, ,
\ee
d$\mathfrak{M}=0$. We also show that the BPS condition appearing as consistency condition for the existence of  supersymmetric purely bosonic solution of the system of mD$0$ equations is expressed by the vanishing of the relative motion Hamiltonian, ${\cal H}_{\text{susy}}=0$, so that the mass of supersymmetric mD$0$ configuration is defined by the above mentioned constant $m$ from the center of mass  part of the action,
\be
\mathfrak{M}_{\text{susy}} = m\; .
\ee

We also find a convenient gauge fixing, which simplifies essentially the relative motion equations, and also
an interesting correspondence  of these with the equations of motion of the maximally supersymmetric $\text{d}=1$ SU($N$) SYM  model. In particular we show that all the supersymmetric solutions of the mD$0$ equations preserve just a half of the spacetime supersymmetry (this is to say are 1/2 BPS states
\footnote{BPS is the abbreviation for Bogomol'nyi--Prasad--Sommerfield.}) and, in its relative motion sector, are in one-to-one correspondence with BPS states of the SYM theory.

The rest of the paper is organized as follows. In Sec.~\ref{sec:mD0action} we describe the action from \cite{Bandos:2022uoz} which possesses the properties expected from the complete nonlinear supersymmetric action of multiple D$0$-brane system so that we call it candidate mD$0$ action or, simplifying terminology, (our)  mD$0$-action. It contains an arbitrary positive definite function $\mathcal{M}(\mathcal{H})$ of the relative motion Hamiltonian $\mathcal{H}$ and in Appendix A we present the explicit form of the $\kappa$-symmetry transformation for any choice of $\mathcal{M}(\mathcal{H})$. In Sec.~\ref{sec:dimRed} we show that a particular representative of this family with $\mathcal{M}(\mathcal{H})$ given in \eqref{cM=m+} can be obtained by dimensional reduction of the action for multiple M0-brane (mM$0$) \cite{Bandos:2012jz} which is described in  Sec.~\ref{sec:11Dvariables}. The dimensional reduction uses essentially  the solution of the constraints for 11D spinor moving frame variables (Lorentz harmonics) in terms of their 10D counterparts given in Eqs. \eqref{hV11=hv10}.  In Sec.~\ref{sec:eom} we vary our mD$0$ action,  obtain the  complete set of equations of motion and study their properties. Particularly, in sec. \ref{sec:eom}.E we describe convenient gauge fixing in which the equations simplify essentially and in \ref{sec:eom}.G establish an interesting  relation of the gauge fixed equations with the equations of maximally supersymmetric 1d SU($N$) SYM theory. Supersymmetric bosonic solutions of our mD$0$ equations are discussed in Sec.~\ref{sec:SUSYbosSol} where it is shown that they obey the BPS equation  $\mathcal{H}= 0$ and, in their relative motion part, are in one-to-one correspondence with  BPS states of the SYM theory.  We conclude and discuss the obtained results in Sec.~\ref{sec:conclusions}. Some technical details can be found in Appendices B and C. In Appendix D we consider  a particular non-supersymmetric solution of mD$0$ equations.

\begin{widetext}

\section{10D mD0 action and its symmetries}
\label{sec:mD0action}

\subsection{mD$0$ fields }

The field content of the mD$p$ system is suggested by its very low energy gauge fixed description given, according to \cite{Witten:1995im}, by
U$(N)$ SYM action (where $N$ is the number of nearly coincident D$p$-branes in the system), and by the known fact that in the case of single D$p$-brane such a description is given by the Abelian U$(1)$ SYM action.
Hence the field content of the Lorentz-covariant formulation of mD$p$-brane system should be given by the fields known from the description of single D$p$-brane, which are essentially coordinate functions and Abelian gauge field, and by the fields of maximally supersymmetric $\text{d}=p+1$ SU$(N)$ SYM multiplet.

In the case of mD$0$-brane the worldline gauge fields, all depending on the proper time $\tau$ which parametrize  the center of mass worldline ${\cal W}^1$,  are coordinate functions
\be
Z^M(\tau)= (x^\mu(\tau),\theta^{\alpha 1}(\tau),\theta_{\alpha}^{2}(\tau)) \; , \qquad \mu=0,...,9\; , \qquad \alpha=1,...,16\; ,
\ee
describing the embedding of the worldline ${\cal W}^1$ in target type IIA superspace $\Sigma^{(10|32)}$ with bosonic and fermionic  coordinates
$Z^M= (x^\mu,\theta^{\alpha 1},\theta_{\alpha}^{2})\; $\footnote{Greek indices from the middle of alphabet, in particular, $\mu=0,...,9$ are spacetime  10-vector indices. In the case of curved target  superspace these should be distinguished  form the target superspace indices which we denote by symbols from the beginning of Latin alphabet, $a=0,...,9$. We find convenient also use both these types of indices in the case of flat target superspace, although in this case they can be identified. The letters from the beginning of Greek alphabet denote 10D Majorana-Weyl spinor indices, e.g. $\alpha=1,...,16$.  }
\be
{\cal W}^1 \; \subset \; \Sigma^{(10|32)} \; : \qquad Z^M= Z^M(\tau)\; ,
\ee
the matrix fields of the $\text{d}=1$ ${\cal N}=16$ SU($N$) SYM multiplet, and some auxiliary fields. The set of these latter includes the spinor moving frame fields, which are described below, and the matrix momentum field which allows to write the SYM action in the first order form.

The set of matrix fields of mD$0$ system includes nine bosonic traceless  Hermitian $N \times N$ matrices ${\bb X}^i$ enumerated by SO$(9)$ vector index $i=1,...,9$, sixteen Hermitian fermionic $N \times N$ matrices  ${\mathbf \Psi}_q$ enumerated by spinor SO(9) (i.e. Spin(9)) index $q=1,...,16$ and anti-Hermitian bosonic traceless $N \times N$ matrix 1-form ${\bb A}=\text{d}\tau {\bb A}_\tau$. In addition we introduce 9-vector Hermitian $N \times N$ field ${\bb P}^i$ which will play a role of momentum conjugate to the matrix field ${\bb X}^i$.

\subsection{Moving frame and spinor moving frame fields (Lorenz harmonics)}

Finally to write the  candidate mD0 action from  \cite{Bandos:2022uoz} we need to introduce the moving frame and spinor moving frame fields.
These are given by SO$(1,9)$ valued and Spin$(1,9)$ valued matrices
\begin{eqnarray}\label{harmV=10D0}
(u_\mu^0, u_\mu^i) \in \text{SO}(1,9)\qquad {\rm and} \qquad v_\alpha{}^q \in \text{Spin}(1,9)\;  \qquad
\end{eqnarray}
which are  related by the conditions of the Lorentz invariance of sigma-matrices
\be\label{us=vsv10D} u^{(b)}_\mu \sigma^\mu_{\alpha\beta}= v_{\alpha}^q \sigma^{(b)}_{qp}v_{\beta}^p\; , \qquad
u^{(b)}_\mu \tilde{\sigma}{}_{(b)}^{qp}= v_{\alpha}^q \tilde{\sigma}{}_{\mu}^{\alpha\beta}v_{\beta}^p\, .
\ee
Here $q, p=1,...,16$ should be identified as spinor indices of SO(9) which is the group of  symmetry preserving the splitting
 of the moving frame matrix in \eqref{harmV=10D0},
\be\label{SO9}
u^{\mu 0}(\tau) \mapsto u^{\mu 0}(\tau) \; , \qquad u^{\mu i}(\tau) \mapsto u^{\mu j}(\tau) {\cal O}^{ji}(\tau)\; , \qquad  {\cal O}^{ji} {\cal O}^{jk}=\delta^{ik}\; .
\ee
This will be one of the gauge symmetries of our model.

Notice that the first of the relations in \eqref{harmV=10D0} implies
\be\label{u02=1}
 u_\mu^0u^{\mu 0}=1\; , \qquad u_\mu^0u^{\mu i}=0\; , \qquad u_\mu^iu^{\mu j}=-\delta^{ij}\; .  \qquad
\ee
Choosing the SO$(9)$ invariant  representation
\be \sigma^{(a)}_{qp}=(\delta_{qp}, \gamma^i_{qp})=\tilde{\sigma}{}_{(a)}^{qp}\; , \qquad \gamma^i_{qp}=\gamma^i_{pq}\; , \qquad \gamma^{(i}\gamma^{j)}=\delta^{ij} \ee in which $\gamma^i_{qp}=\gamma^i_{pq}$ are $\text{d}=9$  gamma matrices, we find that \eqref{us=vsv10D} implies
\begin{eqnarray}\label{u0s=vv}
\sigma^{{\bf 0}}_{\alpha\beta}:=  u_\mu^{{0}} \sigma^\mu_{\alpha\beta}=v_\alpha{}^q v_\beta{}^q \; , \qquad
\sigma^{{\bf i}}_{\alpha\beta}:= u_\mu^{{i}} \sigma^\mu_{\alpha\beta}=v_\alpha{}^q \gamma^i_{qp}v_\beta{}^p \;  \qquad
\\ \nonumber {\rm and}\qquad
v_{\alpha}^q \tilde{\sigma}{}_{\mu}^{\alpha\beta}v_{\beta}^p= u_\mu^{{0}} \delta_{qp}+u_\mu^{{i}} \gamma^i_{qp}\; .  \qquad
\end{eqnarray}
We can also find
\bea
\label{u0ts=vv}
\tilde{\sigma}^{{\bf 0}\alpha\beta}:= u_\mu^0 \tilde{\sigma}^{\mu \alpha\beta}=v_q{}^\alpha v_q{}^\beta \; , \qquad
\tilde{\sigma}^{{\bf i}\alpha\beta}:= u_\mu^i \tilde{\sigma}^{\mu\alpha\beta}=- v_q{}^\alpha  \gamma^i_{qp} v_p{}^\beta\; \qquad
\eea
where
\begin{eqnarray}\label{vs=v-1}
v_\alpha^q \tilde{\sigma}^{{\bf 0}\alpha\beta}=  v_q{}^\beta\; , \qquad  {\sigma}^{{\bf 0}}_{\alpha\beta}v_q{}^\beta  = v_\alpha{}^q\;
\end{eqnarray}
obey
\begin{eqnarray}\label{harmV-1=10D0}
v_q{}^\alpha v_\alpha{}^p=\delta_q{}^p \qquad \Leftrightarrow \qquad
 v_\alpha{}^q v_q{}^\beta= \delta_\alpha{}^\beta
\;  \qquad
\end{eqnarray}
and hence can be identified as inverse spinor frame matrix.

Below we will use the Cartan forms
 \begin{eqnarray}\label{Omi=}
 \Omega^i= u_\mu^0 \text{d}u^{\mu i} , \qquad  \\ \label{Omij=}
  \Omega^{ij}=u_\mu^i \text{d}u^{\mu j}\;  \qquad
\end{eqnarray}
first of which transforms covariantly under SO$(9)$ gauge group \eqref{SO9} while the second has the property of SO$(9)$ connection.
Using \eqref{u02=1} we can express the derivatives of the moving frame variables in terms of the Cartan forms:
 \begin{eqnarray}\label{Du0=}
 \text{D}u_\mu^0:= \text{d}u_\mu^0= u_\mu^i \Omega^i\; , \qquad \text{D}u_\mu^{i}:=\text{d}u_\mu^{i} + u_\mu^{j}\Omega^{ji}=u_\mu^0 \Omega^{i}\; . \qquad
\end{eqnarray}
Here we have also introduced the SO(9) covariant derivatives with the composite connection given by the Cartan form $\Omega^{ji}$ \eqref{Omij=}.

The derivatives of the spinor moving frame matrix is also expressed in terms of the same SO(1,9) Cartan forms
by
\begin{eqnarray}\label{Dv=vOm}
\text{D}v_\alpha{}^q:= \text{d}v_\alpha{}^q+ \frac 1 4 \Omega^{ij} v_\alpha{}^p\gamma^{ij}_{pq}
= \frac 1 2 \gamma^i_{qp} v_\alpha{}^p\Omega^i \; \qquad
\end{eqnarray}
which implies
\begin{eqnarray}\label{Dv-1=vOm}
\text{D}v_q^\alpha &:=& \text{d}v_q^\alpha-  \frac 1 4 \Omega^{ij} \gamma^{ij}_{qp}v_p^\alpha
= -\frac 1 2 v_p^\alpha\gamma^i_{pq} \Omega^i \; . \qquad
\end{eqnarray}

The variation of the moving frame and spinor moving frame variables can be also expressed in terms of Cartan forms or, more precisely, in terms of contraction of these with the variational symbol:
\be
i_\delta \Omega^i = u^{\mu 0}\delta u_\mu^i\; , \qquad i_\delta \Omega^{ij} = u^{\mu i}\delta u_\mu^j\; . \qquad
\ee
The latter parametrize the  SO(9)  transformations which will be the manifest gauge symmetry of our construction so that the essential variations of moving frame vectors and of the spinor moving frame matrix are given by
\bea
&& \delta u_\mu^0= u_\mu^i i_\delta  \Omega^i\; , \qquad \delta u_\mu^{i}=u_\mu^0 i_\delta  \Omega^{i}
\; , \qquad  \\
&& \delta v_\alpha{}^q
= \frac 1 2 \gamma^i_{qp} v_\alpha{}^pi_\delta  \Omega^i\; , \qquad \delta v_q^\alpha  =-\frac 1 2 \gamma^i_{qp}v_p^\alpha i_\delta  \Omega^i~.
\eea

The moving frame formalism allows to define the Lorentz invariant 1-forms on the worldline by contracting the pull-back of the Volkov-Akulov (VA)
1-form of the type IIA superspace
\be\label{Pi=VA}
\Pi^\mu = \text{d}x^\mu -i\text{d}\theta^1\sigma^\mu\theta^1 -i\text{d}\theta^2\tilde{\sigma}^\mu\theta^2 \;  \ee
with moving frame vectors:
\be\label{E0:=}
E^0= \Pi^\mu u_\mu^0 \; , \qquad E^i= \Pi^\mu u_\mu^i \; . \qquad
\ee
The spinor moving frame formalism allows to define on the worldline the following Lorentz invariant fermionic 1-forms:
\be\label{Eq1:=}
E^{1q}= \text{d}\theta^{\alpha 1} v_\alpha^{\; q}\, , \qquad E^{2}_{q}= \text{d}\theta_{\alpha}^{ 2} v_q^\alpha\; . \qquad
\ee

\subsection{Candidate mD$0$ action(s) }

The action for $N$ nearly coincident D$0$-brane (multiple D$0$-brane or mD$0$) system proposed in \cite{Bandos:2022uoz} involves two constants of dimension of mass, $m$ and $\mu$, and has the form
\begin{eqnarray}
\label{SmD0=} && S_{\text{mD0}} = m \int_{\mathcal{W}^1} {E}^{0}     -im  \int_{\mathcal{W}^1} (\text{d}\theta^1\theta^2-\theta^1 \text{d}\theta^2)
     +  \frac 1 {\mu^6} \int_{\mathcal{W}^1}   \left(\text{tr}\left({\bb P}^i \text{D} {\bb X}^i + 4i {\mathbf{ \Psi}}_q \text{D}
{\mathbf{ \Psi}}_q  \right) +  \frac 2 {\cal M} {E}^{0}\, {\cal H}\right) \nonumber \\ && -  \frac 1 {\mu^6} \int_{\mathcal{W}^1} \frac {\text{d} {\cal M}}{ {\cal M} }
 {\rm tr} ({\bb P}^i{\bb X}^i) +   \frac 1 {\mu^6}  \int_{\mathcal{W}^1}
  \frac 1 {\sqrt{2{\cal M}}}({E}{}^{1q}-{E}{}^{2}_{q}) \; {\rm tr}\left(-4i (\gamma^i {\mathbf{ \Psi}})_q  {\bb P}^i  + {1\over 2}
(\gamma^{ij} {\mathbf{ \Psi}})_q  [{\bb X}^i, {\bb X}^j]  \right)  ~.
  \end{eqnarray}
In it ${E}^{0}$ is the projection \eqref{E0:=} of the VA 1-form \eqref{Pi=VA} to the vector $u_\mu^0(\tau)$ of the moving frame
\eqref{harmV=10D0}, ${E}{}^{1q}$ and ${E}{}^{2}_{q}$ are the contractions \eqref{Eq1:=} of the differentials of the first and the second of fermionic coordinate functions with spinor moving frame matrix and with its inverse, respectively. The bosonic
 ${\bb P}^i$ and ${\bb X}^i$ and 16 fermionic ${\mathbf{\Psi}}_q$ are Hermitian traceless $N \times N$ matrix fields and
\begin{eqnarray}
\label{cH=} {\cal H} &=& {1\over 2} \text{tr}\left( {\bb P}^i {\bb P}^i \right)   - {1\over 64}
\text{tr}\left[ {\bb X}^i ,{\bb X}^j \right]^2  - 2\,  \text{tr}\left({\bb X}^i\, {\bf \Psi}\gamma^i {\bf \Psi}\right)  \quad
  \end{eqnarray}
has the meaning of the Hamiltonian of the relative motion of the mD$0$ constituents.
$${\cal M}={\cal M}( {\cal H}/ \mu^6)$$ is an arbitrary nonvanishing function of this Hamiltonian.
Actually the consistency requires the function ${\cal M}={\cal M}( {\cal H}/ \mu^6)$  to be positive definite and we will assume this below,
\be\label{cM>0}
{\cal M}( {\cal H}/ \mu^6) > 0\; .
\ee

The covariant derivatives of matrix fields
\bea\label{DXi=10}
\text{D}\mathbb{X}^i = \text{d}\mathbb{X}^i - \Omega^{ij}\mathbb{X}^j + \left[\mathbb{A},\mathbb{X}^i\right]~,
\\ \label{DPsi=10}
\text{D}\mathbf{\Psi}_q = \text{d}\mathbf{\Psi}_q - \dfrac{1}{4} \Omega^{ij}\gamma^{ij}_{qp}\mathbf{\Psi}_p + \left[\mathbb{A},\mathbf{\Psi}_q\right]~
\eea
include the composite SO(9) connection given by the Cartan form $\Omega^{ij}$ \eqref{Omi=} and the SU$(N)$ connection which is the anti-Hermitian traceless $N \times N$ matrix 1-form  $\mathbb{A}=\text{d} \tau \mathbb{A}_\tau$.

The formal Ricci identities for such a covariant derivatives\footnote{These can be calculated on the extension of  the worldline to some space of two or more dimensions. Such an extension can be realized by considering the forms depending on both differentials and variations of the variables, i.e. on $\text{d}x^\mu(\tau)$ and $\delta x^\mu(\tau)$ etc.}  read
\bea\label{DD=}
\text{DD}{\bb X}^i= \Omega^i\wedge \Omega^j \, {\bb X}^j + [{\bb F}, {\bb X}^i]~, \nonumber
\\
\text{DD}{\mathbf{\Psi}}_q = \frac 1 4 \, \Omega^i\wedge \Omega^j\, (\gamma^{ij}{\mathbf{\Psi}})_q  + [{\bb F}, {\mathbf{\Psi}}_q]~, \label{Ricci}
\eea
where  $\mathbb{F} = \text{d}\mathbb{A} - \mathbb{A} \wedge \mathbb{A}$ is the formal 2-form field strength of the 1-form gauge field $\mathbb{A}$. When deriving \eqref{DD=} we have used the Maurer-Cartan equations
\be\label{MC=Eq}
\text{D} \Omega^i= \text{d}\Omega^i- \Omega^j\wedge \Omega^{ji} = 0 \; , \qquad \text{d}\Omega^{ij}+\Omega^{ik}\wedge \Omega^{kj} =-\Omega^i\wedge \Omega^j\;
\ee
which can be found by taking formal exterior derivatives of \eqref{Du0=}.
These formal expressions  are useful to find the variation of the Lagrangian forms by the method described in Appendix C of \cite{Bandos:2022uoz}.

The action \eqref{SmD0=} is manifestly invariant under 10D type IIA superPoincar\'e transformations, including spacetime (actually target superspace) supersymmetry, as well as under SU$(N)$ gauge symmetry and SO$(9)$ symmetry acting on the suitable indices of moving frame, spinor moving frame and matrix matter fields. Moreover, it is invariant under the worldline supersymmetry transformations the explicit form of which can be found in
\cite{Bandos:2022uoz} as well as in Appendix A. This invariance generalizes $\kappa$-symmetry  of single D$0$-brane action and guarantees that the ground state of the system described by the action \eqref{SmD0=} preserves a part (1/2) of the spacetime supersymmetry.

\section{11D mM0 action and its dimensional reduction to D=10}

In this section we will show that (as was announced in \cite{Bandos:2022uoz}) a particular representative of the family of the candidate mD$0$ actions \eqref{SmD0=}, that with ${\cal M}({\cal H})$ given in \eqref{cM=m+}, can be obtained by dimensional reduction of the 11D mM$0$ action of
\cite{Bandos:2012jz}.

\subsection{11D mM$0$ action and its symmetries}
\label{sec:11Dvariables}

\subsubsection{mM0 center of energy variables}

The mM$0$ brane system which was assumed to be decompactification limit (M-theory lifting) of mD$0$, carries the same matrix fields of $\text{d}=1$ ${\cal N}=16$ SYM on its worldline, but the set of fields describing its center of energy motion is different.  It includes
the coordinate functions
\be
Z^{\underline{M}}(\tau)=  (X^{\underline{\mu}}(\tau), \Theta^{\underline{\alpha}}(\tau))\; , \qquad {\underline{\mu}}=0,1,...,9,10\; , \qquad
{\underline{\alpha}}=1,...,32\;
\ee
which describe the embedding of the worldline in the 11D superspace $\Sigma^{(11|32)}$ with coordinates $Z^{\underline{M}}=  (X^{\underline{\mu}}, \Theta^{\underline{\alpha}})$, the spinor moving frame fields, which we describe below, and the Lagrange multiplier
$\rho^\# (\tau)$, the role of which will be clarified below.

The 11D Volkov-Akulov (VA) 1-form
\be\label{Pi=11}
\Pi^{\underline{\mu}}= \text{d}X^{\underline{\mu}} - i \text{d}\Theta \Gamma^{\underline{\mu}}\Theta = (\Pi^{\mu}, \Pi^{*})
\ee
involves  real symmetric  $32\times 32$ matrices $\Gamma^{\underline{\mu}}_{\underline{\alpha}\underline{\beta}}=\Gamma^{\underline{\mu}}_{\underline{\beta}\underline{\alpha}}= \Gamma^{\underline{\mu}} {}_{\underline{\alpha}}{}^{\underline{\gamma}}C_{\underline{\gamma}\underline{\beta}}$ constructed from 11D Dirac matrices $\Gamma^{\underline{\mu}}{} _{\underline{\alpha}}{}^{\underline{\gamma}}=- (\Gamma^{\underline{\mu}}{}_{\underline{\alpha}}{}^{\underline{\gamma}})^*$  obeying the Clifford algebra \be\label{Cliff}\Gamma^a\Gamma^b+\Gamma^b\Gamma^a = 2\eta^{ab}{\bb I}_{32\times 32}
\ee and the  conjugation matrix $C$ which is   antisymmetric and imaginary in our mostly minus notation, $C_{\underline{\gamma}\underline{\beta}}= - C_{\underline{\beta}\underline{\gamma}}=- (C_{\underline{\gamma}\underline{\beta}})^*$. We will also need the symmetric matrices  with upper indices $\tilde{\Gamma}^{{\underline{\mu}}  \; \underline{\alpha}\underline{\beta}}= \tilde{\Gamma}^{{\underline{\mu}}  \; \underline{\beta}\underline{\alpha}}= C^{\underline{\alpha}\underline{\gamma}} {\Gamma}^{\underline{\mu}}{}_{\underline{\gamma}}{}^{\underline{\beta}}$ which can be used to write the Clifford algebra \eqref{Cliff} in the form  $\Gamma^{(\underline{\mu}}{}_{\underline{\alpha}\underline{\delta}} \tilde{\Gamma}^{\underline{\nu}) \underline{\delta}\underline{\beta}}=\eta^{\underline{\mu}\underline{\nu}} \delta_{\underline{\alpha}}^{\underline{\beta}}$.

We use the following  SO(1,9) invariant decomposition of  11D fermionic Majorana spinor coordinates on two   10D Majorana-Weyl  spinor coordinates of opposite chiralities
\be
\Theta^{\underline{\alpha}}= \left(\begin{matrix} \theta^{1\alpha} \cr
\theta^2_{\alpha}
\end{matrix}\right)\; . \qquad
\ee
Then, with the appropriate $\text{SO}(1,9)$ invariant  representation of the 11D gamma matrices which is presented in Appendix
\ref{AppGamma}, the 11D VA 1-form  splits, as indicated already in \eqref{Pi=11}, in  10D  VA 1-forms \eqref{Pi=VA}
\be\label{Pi=10}
\Pi^{{\mu}}= \text{d}X^{{\mu}} - i\text{d}\theta^1 \sigma^{{\mu}}\theta^1 - i\text{d}\theta^2 \tilde{\sigma}^{{\mu}}\theta^2
\ee
and the scalar 1-form
\be\label{Pi*=}
\Pi^{*}= \text{d}X^{*} + i\text{d}\theta^1 \theta^2 + i \text{d}\theta^2 \theta^1\; .
\ee

The 11D $\text{SO}(1,9)/[\text{SO}(1,1)\times \text{SO}(9)]$ spinor moving frame variables (or 11D Lorentz harmonics \cite{Galperin:1992pz}; see
\cite{Bandos:2012jz,Bandos:2013uoa} and refs therein)  are  defined as rectangular blocks of Spin$(1,10)$-valued matrix
\begin{eqnarray}\label{harmV=11D}
V_{\underline{\alpha}}^{(\underline{\beta})}= \left(\begin{matrix} v_{\underline{\alpha} {q}}^{\; +} , & v_{\underline{\alpha} q}^{\; -}
  \end{matrix}\right) \in \text{Spin}(1,10)\;  \qquad
\end{eqnarray}
which provides a kind of square root of the $\text{SO}(1,10)$ valued moving frame matrix
\begin{eqnarray}\label{uaib=}
U_{\underline{\mu}}^{(\underline{a})}= \left( \frac 1 2 \left(U_{\underline{\mu} }^\#+ U_{\underline{\mu} }^=\right) , U_{\underline{\mu} }^i\, , \frac 1 2 \left(U_{\underline{\mu} }^\# -U_{\underline{\mu}}^=\right)\right)  \in \text{SO}(1, 10)\;  \qquad
\end{eqnarray}
constructed from two light-like vectors normalized in their contraction and 9 orthonormal spacelike vectors orthogonal to that two,
\begin{eqnarray}\label{uu=0}
&& U_{\underline{\mu}}^{=}U^{{\underline{\mu}}=}=0 \; , \qquad
  U_{\underline{\mu}}^{\#}U^{{\underline{\mu}}\#}=0 \; , \qquad  U_{\underline{\mu}}^{=}U^{\underline{\mu}\#}=2 \; , \qquad \\ \label{uui=0} && U_{\underline{\mu}}^{i}U^{{\underline{\mu}}=}=0 \; , \qquad U_{\underline{\mu}}^{i}U^{{\underline{\mu}}\#}=0 \; , \qquad  U_{\underline{\mu}}^{i}U^{{\underline{\mu}}j}=-\delta^{ij} \; . \qquad
\end{eqnarray}
The moving frame vectors also obey $U_{\underline{\mu}}^{(\underline{c})}\eta_{(\underline{c})(\underline{d})}U_{\underline{\nu}}^{(\underline{d})}=\eta_{{\underline{\mu}}{\underline{\nu}}}$, which can be written in the form of
\begin{eqnarray}\label{I=UU}
\delta_{\underline{\mu}}{}^{\underline{\nu}}= {1\over 2}U_{\underline{\mu}}^{=}U^{{\underline{\nu}}\#}+ {1\over 2}U_{\underline{\mu}}^{\#}U^{{\underline{\nu}}=}-  U_{\underline{\mu}}^{i}U^{{\underline{\nu}}i} \; . \qquad
\end{eqnarray}

The above mentioned square root relations have the form of the Lorentz invariance statement for the 11D Dirac matrices
\begin{eqnarray}\label{VGVt=G11} V\Gamma_{\underline{\mu}} V^T =  U_{\underline{\mu}}^{({\underline{a}})} {\Gamma}_{(\underline{a})}\, , \qquad V^T \tilde{\Gamma}^{({\underline{a}})}  V = \tilde{\Gamma}^{{\underline{\mu}}} U_{\underline{\mu}}^{({\underline{a}})}\;
 \, . \qquad \end{eqnarray}
The spinor moving frame variables also obey the constraint
\be \label{VCVt=C11}
 VCV^T=C \;  \qquad
\ee
manifesting the Lorentz invariance of the charge conjugation matrix. This
implies that
 \begin{eqnarray}
\label{V-1=CV-A} v_{q}^{+ {\underline{\alpha}}}  =  i C^{{\underline{\alpha}}{\underline{\beta}}}v_{{\underline{\beta}} q}^{\; +  }   \, ,
\qquad  v_{q}^{- {\underline{\alpha}}}  =  -i C^{{\underline{\alpha}}{\underline{\beta}}}v_{{\underline{\beta}} q}^{\; -  }   \,  \qquad
 \end{eqnarray}
define the inverse spinor moving frame matrix, i.e. that
\begin{eqnarray}\label{v-qv+p=11}
&
v^{-{\underline{\alpha}}}_{{q}}   v_{{\underline{\alpha}} {p}}^{\; +}=\delta_{{q}{p}}
 \; ,  \qquad & v^{-{\underline{\alpha}}}_{{q}}   v_{{\underline{\alpha}} q}^{\; -}=0 \;  , \qquad
 \nonumber  \\
 & v^{+{\underline{\alpha}}}_{{q}}  v_{{\underline{\alpha}} {p}}^{\; +}=0
 \;  , \qquad & v^{+{\underline{\alpha}}}_{{q}} v_{{\underline{\alpha}} {p}}^{\; -} = \delta_{qp} \;  . \qquad
\end{eqnarray}

With an appropriate (SO$(1,1)\times \text{SO}(9)$ invariant) representation of 11D gamma matrices (see Appendix \ref{AppGamma}) Eqs. \eqref{VGVt=G11} split into
\begin{eqnarray}\label{u==v-v-=11D}
 &
  v^-_{{q}} \tilde{\Gamma}_{{\underline{\mu}}}v^-_{{p}}=U_{\underline{\mu}}^= \delta_{{q}{p}}   \; , & \qquad  U_{\underline{\mu}}^= \Gamma^{\underline{\mu}}_{\underline{\alpha}\underline{\beta}}= 2v_{\underline{\alpha} q}{}^- v_{\underline{\beta} q}{}^-  \; ,  \qquad \\
\label{v+v+=u++}
&  \;  v_{{q}}^+ \tilde{\Gamma}_{{\underline{\mu}}} v_{{p}}^+ =U_{{\underline{\mu}}}^{\# } \delta_{{q}{p}}\; , & \qquad U_{ {\underline{\mu}}}^{\# } {\Gamma}^{ {\underline{\mu}}}_{ {{\underline{\alpha}}} {{\underline{\beta}}}} =2 v_{{{\underline{\alpha}}}{q}}{}^{+}v_{{{\underline{\beta}}}{q}}{}^{+} \; , \qquad \\
 \label{uIs=v+v-=11D}
& v_{{q}}^- \tilde {\Gamma}_{{\underline{\mu}}} v_{{p}}^+=U_{ {\underline{\mu}}}^{i} \gamma^i_{q{p}}\; , &\qquad
 U_{{\underline{\mu}}}^{i} {\Gamma}^{{\underline{\mu}}}_{{\underline{\alpha}}{\underline{\beta}}} =2 v_{( {{\underline{\alpha}}}|{q} }{}^- \gamma^i_{qp}v_{|{{\underline{\beta}}}){p}}{}^{+} \; , \quad  \end{eqnarray}
where $\gamma^i_{qp}=\gamma^i_{p{q}}$ are SO(9) Dirac matrices, \be \gamma^{i}_{qr}\gamma^{j}_{rp} +\gamma^{j}_{qr}\gamma^{i}_{rp} =2\delta^{ij}\delta_{qp}\; . \ee

The derivative of the moving frame and spinor moving frame variables can be expressed in terms of Cartan forms
\bea\label{Omij=11D}\underline{\Omega}^{ij}:=  U^{\underline{\mu}i}\text{d} U_{\underline{\mu}}^j\, ,  \qquad \\ \label{Om0=11D} \underline{\Omega}^{(0)}:=  U^{\underline{\mu}=}\text{d} U_{\underline{\mu}}^{\#}\, ,  \qquad  \\ \label{Om--j=11D} \underline{\Omega}^{\#j}:=  U^{\underline{\mu}\#}\text{d} U_{\underline{\mu}}^j\, , \qquad \underline{\Omega}^{=j}:=  U^{\underline{\mu}=}\text{d} U_{\underline{\mu}}^j\,  \qquad
\eea (see e.g. \cite{Bandos:2017zap,Bandos:2017eof} and refs. therein for more details).
The form (\ref{Omij=11D}) transforms as connection under the SO$(9)$  symmetry acting on the 9-vector indices $i,j$ and 16 component spinor indices
$q,p$ of the moving frame variables and matrix fields. This is the counterpart of the
Cartan form \eqref{Omij=} of the 10D SO(1,9)/SO(9)  spinor moving frame (Lorentz harmonics) formalism.

The form (\ref{Om0=11D}) transforms as the connection under the SO$(1,1)$ group acting on
the moving frame, spinor moving frame variables and Lagrangian multiplier $\rho^\#$ according to the weights indicated by their sign indices.
In particular  $\#=++$ so that under SO$(1,1)$
\bea\label{SO11}
U^\#_{\underline{\mu}} \mapsto e^{2\beta } U^\#_{\underline{\mu}}\; , \qquad U^=_{\underline{\mu}} \mapsto e^{-2\beta } U^=_{\underline{\mu}}\; , \qquad U^i_{\underline{\mu}} \mapsto  U^i_{\underline{\mu}}\; , \qquad \nonumber  \\
\underline{\Omega}^{(0)}  \mapsto   \underline{\Omega}^{(0)} +d\beta   \; , \qquad \underline{\Omega}^{ij}  \mapsto   \underline{\Omega}^{ij} \; , \qquad  \nonumber \\
\rho^\#\mapsto e^{2\beta }\rho^\#\; . \qquad
\eea
The  forms \eqref{Om--j=11D} are covariant with respect to both symmetries which will serve as gauge symmetry of the mM$0$ action.

\subsubsection{Matrix fields describing the relative motion of mM0 constituents}

Matrix fields describing the relative motion of the mM0 constituents are exactly the same as those of the mD$0$-system: the bosonic traceless $N \times N$ matrices ${\bb X}^i$ and ${\bb P}^i$, carrying the SO(9) vector indices,  fermionic traceless $N \times N$ matrices ${\bf \Psi}_q$ carrying SO(9) spinor index $q=1,..,16$ and bosonic traceless $N \times N$ matrix 1-form ${\bb A}= \text{d}\tau {\bb A}_\tau$.

\subsubsection{mM0 action}

The action for the 11D mM0 system proposed in \cite{Bandos:2012jz} can be written in the form
\begin{eqnarray}
\label{SmM0=} && S_{\text{mM}0} = \int_{\mathcal{W}^1} \rho^{\#}\, \underline{E}^{=} +  \frac 1 {\mu^6} \int_{\mathcal{W}^1}  \left(  \text{tr}\left({\bb P}^i \text{D} {\bb X}^i + 4i \mathbf{ \Psi}_q \text{D}
\mathbf{\Psi}_q  \right) + \underline{E}^{\#} \,\frac 1 {\rho^{\#}}\, {\cal H} \right)+ \quad \nonumber \\ && +   \frac 1 {\mu^6}  \int_{\mathcal{W}^1}
  \underline{E}{}^{+q} \,\frac 1 {\sqrt{\rho^{\#}}}\, \text{tr}\left(-4i (\gamma^i \mathbf{\Psi})_q  {\bb P}^i + {1\over 2}
(\gamma^{ij} \mathbf{\Psi})_q  [{\bb X}^i, {\bb X}^j]  \right) -  \frac 1 {\mu^6} \int_{\mathcal{W}^1} \frac {\text{D}\rho^\# }{\rho^\# }
\, \text{tr} ({\bb P}^i{\bb X}^i)~, \qquad
  \end{eqnarray}
in which the matrix fields are redefined (with respect to the ones used in \cite{Bandos:2012jz}) to be inert under the SO$(1,1)$ symmetry acting on spinor moving frame variables and Lagrange multiplier
$\rho^\#$ \eqref{SO11}.

The covariant derivatives of the matrix fields in \eqref{SmM0=}
 \begin{eqnarray}
\label{DXi=11D} \text{D}{\bb X}^i  &:=& \text{d}{\bb X}^i   - \underline{\Omega}^{ij} {\bb
X}^j+ [{\bb A},    {\bb X}^i] \; , \qquad \\ \label{DPsi=11D} \text{D}\mathbf{\Psi}_q  &:=& \text{d} \mathbf{\Psi}_q
   -{1\over 4} \underline{\Omega}^{ij} \gamma^{ij}_{qp} \mathbf{\Psi}_p+ [{\bb A},
 \mathbf{\Psi}_q ] \;  \qquad
  \end{eqnarray}
 include, besides the SU($N$) connection 1-form ${\bb A}$, also the composite SO(9) connection  $\underline{\Omega}^{ij}$ \eqref{Omij=11D},
 while
  \be
  \text{D}\rho^\#= \text{d}\rho^\#- 2\underline{\Omega}^{(0)} \rho^\#\;
  \ee
includes the SO(1,1) connection \eqref{Om0=11D}. Let us recall that these composite connections are Cartan forms representing the nonvanishing components of the derivatives of the moving frame and spinor moving frame variables.

The moving frame and spinor moving frame variables enter the action also explicitly through the projections
\be\label{E++=}
\underline{E}^\#=\Pi^{\underline{\mu}} U_{\underline{\mu}}^\# \; , \qquad \underline{E}^==\Pi^{\underline{\mu}} U_{\underline{\mu}}^= \;  \qquad
\ee
of the VA 1-form \eqref{Pi=11}, which are the bosonic supervielbein forms of flat 11D superspace,   and
\be\label{E+q=}
\underline{E}^{+q} = \text{d}\Theta^{\underline{\alpha}}\, v^{\; +}_{\underline{\alpha} q}\,
\ee
of the pull-back of the fermionic supervielbein form of this superspace.

The relative motion Hamiltonian in \eqref{SmM0=}
\begin{eqnarray}
\label{HmM0=} {\cal H} &=& {1\over 2} \text{tr}\left( {\bb P}^i {\bb P}^i \right) - {1\over 64}
\text{tr}\left[ {\bb X}^i ,{\bb X}^j \right]^2  - 2\,  \text{tr}\left({\bb X}^i\, \mathbf{\Psi}\gamma^i\mathbf{\Psi}\right)  \qquad
  \end{eqnarray}
coincides with that of the mD0 system, Eq. \eqref{cH=}.

The action \eqref{SmM0=} has the manifest 11D target superspace supersymmetry and, as it was shown in \cite{Bandos:2012jz}, is invariant under 16 parametric worldline supersymmetry generalizing the $\kappa$-symmetry  of single M$0$-brane action in  its spinor moving frame formulation of \cite{Bandos:2007mi,Bandos:2007wm}.

The properties of this mM$0$ system described by the action \eqref{SmM0=} were further studies in
\cite{Bandos:2013uoa}. The problem of its dimensional reduction to $\text{D}=10$ was addressed in \cite{Bandos:2018ntt} but was not solved there. The reason was that a suitable convenient choice of
the basic matrix fields was not found in \cite{Bandos:2018ntt}; we have found it more  recently, first for the simplified case of $\text{D}=4$ counterpart of mM$0$ system \cite{Bandos:2021vrq}, and we have already written the
mM$0$ action of \cite{Bandos:2012jz} in terms of these fields in Eq. \eqref{SmM0=}. The dimensional reduction
of this mM$0$ brane action is the subject of the next (sub)section.

\subsection{Dimensional reduction of mM$0$ action to $\text{D}=\textbf{10}$}
\label{sec:dimRed}

As we have already noticed, with the suitable representation for 11D Gamma matrices and charge conjugation matrix  which can be found in Appendix \ref{AppGamma} (see Eqs.  (\eqref{G11=s10}, \eqref{tG11=s10} and \eqref{C11=10}) the 11D Volkov-Akulov (VA) 1-form is split as in \eqref{Pi=11}
into 10D  VA 1-form \eqref{Pi=10} and SO(1,9) invariant 1-form \eqref{Pi*=}.

\subsubsection{SO(1,9) invariant expressions for 11D spinor moving frame and  basic 1-forms}

The next stage is to solve the  constraints defining the above described 11D  spinor moving frame variables in terms of   Spin$(1,9)/\text{Spin}(9)$ spinor moving frame variables (\ref{harmV=10D0}).
The convenient form of the solution is
\begin{eqnarray}\label{hV11=hv10}
\frac 1 {\sqrt{\rho^\#}}  v_{{\underline{\alpha}} q}^{\; +}
=\frac 1 {\sqrt{2}}  \frac 1 {\sqrt{{\cal M}}}\left(\begin{matrix} v_{{\alpha}}^{\; q} \cr  -v_q{}^\alpha
  \end{matrix}\right)\; , \qquad  \sqrt{\rho^\#} \,v_{{\underline{\alpha}} q}^{\; -}
=\frac 1 {\sqrt{2}}\, \sqrt{{\cal M}}\,  \left(\begin{matrix} v_{{\alpha}}^{\; q} \cr v_q{}^\alpha
  \end{matrix}\right)\;  \qquad
\end{eqnarray}
which implies the complementary relations
 \begin{eqnarray}\label{hV-1=hv10}
 \frac 1 {\sqrt{\rho^\#}}  v^{+\underline{\alpha}}_q
=\frac 1 {\sqrt{2}} \frac 1 {\sqrt{{\cal M}}}  \left(\begin{matrix}  v_q{}^\alpha \cr v_{{\alpha}}^{\; q}
  \end{matrix}\right)\; , \qquad   \sqrt{\rho^\#} \, v^{-\underline{\alpha}}_q
=\frac 1 {\sqrt{2}} \, \sqrt{{\cal M}}\, \left(\begin{matrix}  v_q{}^\alpha \cr -v_{{\alpha}}^{\; q}
  \end{matrix}\right)\; .   \qquad
\end{eqnarray}
Eqs. \eqref{hV11=hv10} involve the Lagrange multiplier $\rho^\#$ of the mM$0$ action and also an arbitrary function ${\cal M}={\cal M}(\tau)$ of the proper  time parametrizing the center of energy  worldline of the mM$0$ system.

Let us stress that this is not an ansatz but rather a general solution of the constraints \eqref{u==v-v-=11D}--\eqref{uIs=v+v-=11D} and \eqref{VCVt=C11}. This can be easily checked by counting the number of degrees of freedom, modulo natural gauge symmetries, in the left and right hand sides of the relation
\eqref{hV11=hv10} which gives  1+$9=1$+9. Indeed, both sides include scalar functions, $\rho^\#$ and ${\cal M}$ respectively, and spinor frame variables parametrizing cosets isomorphic to ${\bb S}^9$ sphere: $\frac {\text{SO}(1,10)}{(\text{SO}(1,1)\times \text{SO}(9)) \subset\!\!\!\!\times K_9}\simeq {\bb S}^9$ and $\frac {\text{SO}(1,9)}{\text{SO}(9)}\simeq {\bb S}^9$, respectively (see e.g. \cite{Bandos:2017zap, Bandos:2017eof} and refs. therein for more details). Notice also that the l.h.s. of \eqref{hV11=hv10} preserves the characteristic $\text{SO}(1,1)$ gauge symmetry \eqref{SO11} of the 11D spinor moving frame formalism (which was taken into account in the above counting of the degrees of freedom).

As far as the 11D Lorentz symmetry $\text{SO}(1,10)$ is concerned, the expressions \eqref{hV11=hv10} break this down to its $\text{SO}(1,9)$ subgroup which becomes 10D Lorentz symmetry of the reduced theory.

Using \eqref{hV11=hv10} we obtain  the following  expression for (the pull-backs of) 11D fermionic supervielbein forms which enter the action \eqref{SmM0=}
\be  \underline{E}{}^{+q} = {\sqrt{\rho^{\#}}}\, \frac 1 {\sqrt{2{\cal M}}} (E^{1q} - E_q^2) \; , \ee
where
\be\label{E1q=} E^{1q}=\text{d}\theta^{1\alpha} v_{\alpha}^q \; , \qquad E^2_{q}=\text{d}\theta^2_{\alpha} v_q^{\alpha}\;  \qquad \ee
can be naturally identified with the pull-backs of the fermionic $\text{D}=10$ type IIA supervielbein forms \eqref{Eq1:=}.

Now, substituting \eqref{hV11=hv10} into Eqs. \eqref{u==v-v-=11D} and \eqref{v+v+=u++}
and using the suitable representation for 11D gamma matrices (see Appendix \ref{AppGamma}) as well as
\eqref{u0s=vv} we find that the 11D moving frame vectors are related to its 10D counterparts by
\bea\label{U==}
U^=_{\mu}= \frac {{\cal M}} {\rho^\#} u^0_{\mu}\; , \qquad U^=_{*}= - \frac {{\cal M}} {\rho^\#}\; , \qquad
\\ \label{U++=}
U^\#_{\mu} =  \frac {\rho^\#}{{\cal M}} \, u^0_{\mu} \; , \qquad U^\#_{*}  = \frac {\rho^\#}{{\cal M}}\;  \qquad
\eea
so that
\be
\rho^\#\underline{E}^{=}= {\cal M}({\rm E}^0 -\Pi^*) \; , \qquad \frac {\underline{E}^{\#}}{\rho^\#}= \frac 1 {{\cal M}}({\rm E}^0 +\Pi^*) \; .  \qquad
\ee
In the same way we find from \eqref{uIs=v+v-=11D} \be\label{Ui=ui} U_{\underline{\mu}}^i = \delta_{\underline{\mu}}{}^\nu u_\nu^i\;
\ee
which can be used to check that  the composite $\text{SO}(9)$ connection of the  mM$0$ system coincides with its 10D counterpart \eqref{Omij=},
\be\label{Omij11=10}\underline{\Omega}^{ij}:=  U^{\underline{\mu}i}\text{d} U_{\underline{\mu}}^j = u^{{\mu}i}\text{d} u_{{\mu}}^j=: {\Omega}^{ij}\; .
\ee
Thus the covariant derivative of the matrix fields in the mM$0$ action \eqref{SmM0=}, Eqs. \eqref{DXi=11D} and  \eqref{DPsi=11D}, coincide with \eqref{DXi=10} and \eqref{DPsi=10} used in the mD$0$ action \eqref{SmD0=}.

As far as $\text{SO}(1,1)$ connection is concerned, we find from \eqref{U==} and \eqref{U++=}
\be \underline{\Omega}{}^{(0)}= \frac 1 4 U^{=\underline{\mu}}\text{d}U_{\underline{\mu}}^\# = \frac 1 2 \left( \frac {\text{d}\rho^{\#}}{\rho^{\#}}- \frac {\text{d}{\cal M}}{{\cal M}}\right)\; .   \ee
In particular, this implies that
\be
\text{D}\rho^\# = \text{d}\rho^\#- 2 \rho^\# \underline{\Omega}^{(0)} = \rho^\# \frac {\text{d}{\cal M}}{{\cal M}}\; .
\ee
At this stage one can observe that, after using in \eqref{SmM0=} the relations \eqref{hV11=hv10} and their consequences, $\rho^\#$ disappears from the  11D  action being replaced by ${\cal M}$.

\subsubsection{Dimensional reduction of mM0 action }

Thus with the solution \eqref{hV11=hv10} and the splitting \eqref{Pi=11} the mM$0$ action \eqref{SmM0=} acquires the form
\begin{eqnarray}
\label{SmM0=V11-v10}  S_{\text{mM}0}\vert_{\eqref{hV11=hv10}} &=&  \int_{\mathcal{W}^1} {\cal M} ({E}^{0}-\Pi^*)  +  \frac 1 {\mu^6} \int_{\mathcal{W}^1} \frac 1 {\cal M} ({E}^{0}+\Pi^*)\, {\cal H}  + \quad \nonumber \\ && +  \frac 1 {\mu^6} \int_{\mathcal{W}^1}    \text{tr}\left({\bb P}^i \text{D} {\bb X}^i + 4i \mathbf{\Psi}_q \text{D}
\mathbf{\Psi}_q  \right)-  \frac 1 {\mu^6} \int_{\mathcal{W}^1} \frac {\text{d} {\cal M}}{ {\cal M} }
\, \text{tr} ({\bb P}^i{\bb X}^i) + \quad \nonumber \\ && +   \frac 1 {\mu^6}  \int_{\mathcal{W}^1}
  \frac 1 {2\sqrt{{\cal M}}}({E}{}^{1q}-{E}{}^{2}_{q}) \, \text{tr}\left(-4i (\gamma^i \mathbf{\Psi})_q  {\bb P}^i + {1\over 2}
(\gamma^{ij} \mathbf{\Psi})_q  [{\bb X}^i, {\bb X}^j]  \right) , \qquad
  \end{eqnarray}
where the covariant derivatives are defined in \eqref{DXi=10}, \eqref{DPsi=10}, the relative motion Hamiltonian  ${\cal H}$ has the form of \eqref{cH=},
\be
E^0=\Pi^\mu u_\mu^0
\ee
with $\Pi^\mu$ from \eqref{Pi=10} (see  \eqref{E0:=}), $\Pi^*$ is given in \eqref{Pi*=} and, according to \eqref{E1q=},
\be\label{E1q-2=} E^{1q}-E^2_{q} = \text{d}\theta^{1\alpha} v_{\alpha}^q - \text{d}\theta^2_{\alpha} v_q^{\alpha}\; .  \qquad \ee

The dimensional reduction is then completed by deriving  from the action \eqref{SmM0=V11-v10} the equation of motion for eleventh  bosonic coordinate field $X^*$,
\be\label{cMEq=}
\text{d}\left({\cal M} - \frac 1 {{\cal M}}\, \frac {\cal H} {\mu^6}\right)=0\;
\ee
and substituting its solution back into the functional \eqref{SmM0=V11-v10}.

Eq. \eqref{cMEq=} can be equivalently written in the form
\be\label{cMEq=m}
{\cal M} - \frac 1 {{\cal M}}\, \frac {\cal H} {\mu^6}=m \;
\ee
with some constant $m$. Its solution which has a nonvanishing  limit when ${\cal H}\mapsto 0$ is
\be\label{cM=m+s}
{\cal M} = \frac m 2 +  \sqrt{\frac {m^2} {4}+\frac {\cal H} {\mu^6}}\; .
\ee
Substituting the result back into \eqref{SmM0=V11-v10} we find the action \eqref{SmD0=}  with a particular function
$
{\cal M}= {\cal M} \left( {{\cal H}} /{\mu^6} \right)
$
 given by \eqref{cM=m+s}.

Thus just this particular case of \eqref{SmD0=} model has the apparent 11D origin.

\section{Equations of motion of 10D mD$0$ }
\label{sec:eom}

In this section we write the complete set of equations of motion for the multiple D0-brane system which follow from our action \eqref{SmD0=}.

\subsection{Equations for the center of energy variables}

Beginning from the center of energy variables, we re-group  the variation with respect to the coordinate  functions, $\delta Z^M(\tau)=( \delta x^\mu(\tau), \delta \theta^{\alpha 1}(\tau), \delta \theta^2_{\alpha}(\tau))$, into
\be i_\delta E^i = \delta Z^M(\tau) E_M^a (Z^M(\tau))u_a^i(\tau)\; , \qquad i_\delta E^0 = \delta Z^M(\tau) E_M^a (Z^M(\tau))u_a^0(\tau)\; \qquad
\ee
and
\be
i_\delta (E^{q1} - E_q^2) \; , \qquad i_\delta (E^{q1} + E_q^2) \; , \qquad
\ee
where
\be
i_\delta E^{q1} = \delta \theta^{\alpha 1}v_\alpha{}^q \; , \qquad
i_\delta E_q^{2} = \delta \theta^2_{\alpha}v_q{}^\alpha\; . \qquad
\ee
This choice simplifies the form of the equations of  motion for the coordinate functions (originally defined as $\frac{\delta S_{\text{mD}0}}{\delta Z^M(\tau)}=0$) and splits their set in a Lorentz-covariant manner into
\bea\label{Omi()=0}
\Omega^i \left(m + \frac{2}{\mu^6}\frac{\mathcal{H}}{\mathcal{M}} \right) = 0\; , \qquad \\ \label{E1+E2=Omi}
\left(m + \frac{1}{\mu^6} \frac{\mathcal{H}}{\mathcal{M}} \right)\left(E^{1q} + E^2_{q} \right) = \frac{-i}{4 \sqrt{2\mathcal{M}} \mu^6}\gamma^i_{qp}i\nu_p\Omega^i\; ,
\eea
 and
\bea\label{dH()=0}
\frac{2}{\mathcal{M}} \left(1 - \frac{1}{\mu^6}\frac{\mathcal{M'}}{\mathcal{M}} \mathcal{H} \right)\text{d}\mathcal{H}=0\; , \qquad \\
\label{eq:Dnu}
\frac{1}{\sqrt{2\mathcal{M}}}i\text{D}\nu_q - \frac{1}{\mu^6} \frac{1}{2\sqrt{2 \mathcal{M}}} \frac{\mathcal{M'}}{\mathcal{M}} i \nu_q \text{d} \mathcal{H} - \frac{2i}{\mathcal{M}} \left(E^{1q} - E^2_{q} \right) \mathcal{H} = 0\; ,
\eea
where
\be\label{inu=}
i\nu_q:= {\rm tr} \left(-4i (\gamma^i {\bf{\Psi}})_q  {\bb P}^i + {1\over 2}
(\gamma^{ij} {\bf{\Psi}})_q  [{\bb X}^i, {\bb X}^j]  \right)\; . \qquad
\ee

As we will show below, Eqs. \eqref{dH()=0} and \eqref{eq:Dnu}  are dependent, this is to say they are satisfied identically when other equations are taken into account. This statement is the content of Noether identities for
the worldline  reparametrization gauge symmetry and for the worldline supersymmetry ($\kappa$-symmetry) the ``local parameters'' of which can be identified with
$i_\delta E^0$ and $i_\delta (E^{q1}-E_q^2)$, respectively. Keeping this in mind, we will first discuss \eqref{Omi()=0}, \eqref{E1+E2=Omi} and the equations for spinor frame variables and then turn to the matrix equations (which will then allow to make the above  statements about dependence of
 \eqref{dH()=0} and \eqref{eq:Dnu}).

As far as the function ${\cal M}(\mathcal{H} / \mu^6)$ in our action is positive definite, \eqref{cM>0},
Eq. \eqref{Omi()=0}  implies
\be\label{Omi=0}
\Omega^i = 0 \, . \ee
Taking this into account, we find that  the fermionic equations \eqref{E1+E2=Omi} simplifies to
\be\label{E1+E2=0}
E^{1q} + E^2_{q}  =0  \; .
\ee
Finally, the essential variations of the moving frame and spinor moving frame variables,
\be
\delta u_\mu^0= u_\mu^i i_\delta \Omega^i\, , \qquad  \delta v_\alpha{}^q=  \frac 1 2 \, i_\delta \Omega^i\, \gamma^i_{qp}  v_\alpha{}^p\, , \qquad \delta v_q{}^\alpha=-  \frac 1 2 \, i_\delta \Omega^i\, v_p{}^\alpha \gamma^i_{pq}  \, , \qquad
\ee
result in
\begin{equation}\label{Ei()=ff+Om}
E^i \left(m + \frac{2}{\mu^6} \frac{\mathcal{H}}{\mathcal{M}} \right) - \frac{1}{2 \mu^6  \sqrt{2 \mathcal{M}}}\left(E^{1q} + E^2_{q} \right) \gamma^{i}_{qp}i\nu_p - \frac 2 {\mu^6 } \, \text{tr} \left( \mathbb{P}^{\left[i \right.} \mathbb{X}^{\left. j \right]} + i \mathbf{\Psi} \gamma^{ij} \mathbf{\Psi} \right)\, \Omega^j = 0\; ,
\end{equation}
where $i\nu_q$ is defined in  \eqref{inu=}.  Taking into account  \eqref{Omi=0} and the fermionic equations \eqref{E1+E2=0}, we see that   \eqref{Ei()=ff+Om} finally simplifies to
\begin{equation}\label{Ei=0}
 E^i  = 0\; .
\end{equation}

Thus the set of equations for the center of energy variables  is given by
Eqs. \eqref{Omi=0}, \eqref{E1+E2=0} and \eqref{Ei=0}. This coincides with the set of equations of motion for single D0-brane in its spinor moving frame formulation \cite{Bandos:2000tg}.

\subsection{Equations of the relative motion of the constituents of mD$0$ system}
The equations of motion for the matrix fields describing the relative motion of the mD$0$ constituents read
\bea\label{DXi=}
\text{D}\mathbb{X}^i &=& -\frac{2}{\mathcal{M}} \left(1 - \frac{1}{\mu^6}\frac{\mathcal{M'}}{\mathcal{M}}\mathcal{H} \right)E^0 \mathbb{P}^i + \frac{1}{\mu^6}\frac{\mathcal{M'}}{\mathcal{M}} \left( \mathbb{X}^i  \text{d}\mathcal{H} - \mathbb{P}^i  \text{d}\mathcal{K}\right) + \frac{1}{\sqrt{2\mathcal{M}}} \left(E^{1q} - E^2_{q} \right) \left(4i \left(\gamma^i \mathbf{\Psi} \right)_q - \frac{1}{2\mu^6}\frac{\mathcal{M'}}{\mathcal{M}} i \nu_q \mathbb{P}^i \right)\; , \qquad
\\ \label{DPi=}
\text{D} \mathbb{P}^i &=& \frac{2}{\mathcal{M}} \left[ \left(1- \frac{1}{\mu^6} \frac{\mathcal{M'}}{\mathcal{M}}\mathcal{H} \right)E^0 + \frac{1}{\mu^6}\frac{\mathcal{M'}}{4 \sqrt{2 \mathcal{M}}} \left(E^{1q} - E^2_{q} \right) i\nu_q \right] \left(\frac{1}{16} \left[[\mathbb{X}^i, \mathbb{X}^j],\mathbb{X}^j  \right] - \gamma^i_{pr} \left\lbrace\mathbf{\Psi}_p, \mathbf{\Psi}_r \right\rbrace  \right)
\nonumber \\
& &- \frac{1}{\sqrt{2 \mathcal{M}}}\left(E^{1q}- E^2_{q}\right) [\left(\gamma^{ij} \mathbf{\Psi} \right)_q, \mathbb{X}^j]
 + \frac{1}{\mu^6} \frac{\mathcal{M'}}{\mathcal{M}}\text{d}\mathcal{K} \left(\frac{1}{16} \left[[\mathbb{X}^i, \mathbb{X}^j],\mathbb{X}^j  \right] - \gamma^i_{pr} \left\lbrace\mathbf{\Psi}_p, \mathbf{\Psi}_r \right\rbrace  \right) - \frac{1}{\mu^6}\frac{\mathcal{M'}}{\mathcal{M}} \mathbb{P}^i \text{d}\mathcal{H}\; , \qquad \\
\label{DPsi=} \text{D}\mathbf{\Psi}_q &=& - \frac{1}{2 \sqrt{2\mathcal{M}}}\left(E^{1p}- E^2_{p} \right) \left(\gamma^i_{pq}\mathbb{P}^i + \frac{i}{8}\gamma^{ij}_{pq}[\mathbb{X}^i,\mathbb{X}^j] \right) \nonumber  \\ && - \frac{i}{2\mathcal{M}} \left( \left(1 - \frac{1}{\mu^6} \frac{\mathcal{M'}}{\mathcal{M}}\mathcal{H}\right)E^0 +  \frac{\mathcal{M'}}{2\mu^6} \text{d}\mathcal{K}   + \frac{1}{4 \mu^6}\frac{\mathcal{M'}}{\sqrt{2\mathcal{M}}}\left(E^{1p}- E^2_{p} \right) i\nu_p  \right) [(\gamma^i \mathbf{\Psi})_q, \mathbb{X}^i]  \; ,
\eea
and
\begin{equation}\label{Gauss=}
[\mathbb{X}^i, \mathbb{P}^i] = 4i \left\lbrace \mathbf{\Psi}_q, \mathbf{\Psi}_q \right\rbrace~.
\end{equation}
In  Eqs. \eqref{DXi=}--\eqref{DPsi=}
 ${\cal H}$ is given in \eqref{cH=},
\be\label{cK=}
 {\cal K}= {\rm tr} ({\bb X}^i\,{\bb P}^i)\;
\ee
and $i\nu_q$ is defined in \eqref{inu=}.
Eq. \eqref{Gauss=} appears as a result of variation with respect to the   worldline gauge field $\mathbb{A} = \text{d}\tau \mathbb{A}_\tau$ and is the (non-Abelian version of the)  Gauss law of 10D SYM  reduced to 1d.

Using Eqs. \eqref{DXi=}--\eqref{DPsi=} and \eqref{Gauss=}   one can check (by straightforward although a bit involving calculations) that
\be\label{dH=0}
\text{d}{\cal H}=0
\ee
and
\be\label{Dnu=EH}
i\text{D}\nu_q= \frac {2\sqrt{2}}{\sqrt{{\cal M}}}(E^{1q}-E_q^2){\cal H}\; ,
\ee
which implies that  Eqs. \eqref{dH()=0} and \eqref{eq:Dnu} are satisfied identically (as we have announced and discussed just after their derivation).

Taking into account \eqref{dH=0} we find that the equations for bosonic  matrix matter fields, \eqref{DXi=} and \eqref{DPi=} simplify a bit:
\bea\label{DXi==}
\text{D}\mathbb{X}^i &=& -\frac{2}{\mathcal{M}} \left(1 - \frac{1}{\mu^6}\frac{\mathcal{M'}}{\mathcal{M}}\mathcal{H} \right)E^0 \mathbb{P}^i - \frac{1}{\mu^6}\frac{\mathcal{M'}}{\mathcal{M}} \mathbb{P}^i  \text{d}\mathcal{K} + \frac{1}{\sqrt{2\mathcal{M}}} \left(E^{1q} - E^2_{q} \right) \left(4i \left(\gamma^i \mathbf{\Psi} \right)_q - \frac{1}{2\mu^6}\frac{\mathcal{M'}}{\mathcal{M}} i \nu_q \mathbb{P}^i \right)\; , \qquad
\\ \label{DPi==}
\text{D} \mathbb{P}^i &=& \frac{2}{\mathcal{M}} \left[ \left(1- \frac{1}{\mu^6} \frac{\mathcal{M'}}{\mathcal{M}}\mathcal{H} \right)E^0 + \frac{1}{\mu^6}\frac{\mathcal{M'}}{4 \sqrt{2 \mathcal{M}}} \left(E^{1q} - E^2_{q} \right) i\nu_q \right] \left(\frac{1}{16} \left[[\mathbb{X}^i, \mathbb{X}^j],\mathbb{X}^j  \right] - \gamma^i_{pr} \left\lbrace\mathbf{\Psi}_p, \mathbf{\Psi}_r \right\rbrace  \right)
\nonumber \\
& &- \frac{1}{\sqrt{2 \mathcal{M}}}\left(E^{1q}- E^2_{q}\right) [\left(\gamma^{ij} \mathbf{\Psi} \right)_q, \mathbb{X}^j]
 + \frac{1}{\mu^6} \frac{\mathcal{M'}}{\mathcal{M}}\text{d}\mathcal{K} \left(\frac{1}{16} \left[[\mathbb{X}^i, \mathbb{X}^j],\mathbb{X}^j  \right] - \gamma^i_{pr} \left\lbrace\mathbf{\Psi}_p, \mathbf{\Psi}_r \right\rbrace  \right) \; . \qquad
\eea

This form of equations is not yet final since it involves $\text{d}\mathcal{K}={\rm tr} \left(
\text{D} \mathbb{X}^i \, \mathbb{P}^i+\mathbb{X}^i\text{D} \mathbb{P}^i
\right) $. Calculating the r.h.s. of this expression with the use of \eqref{DXi==} and \eqref{DPi==}, we find
\be\label{dcK==}
\text{d}\mathcal{K}= - \frac 2 {{\cal M}}  \frac {\mathfrak{H} }{ \left(1 + \frac{1}{\mu^6} \frac{\mathcal{M'}}{\mathcal{M}}\mathfrak{H} \right)}\,  \left(1- \frac{1}{\mu^6} \frac{\mathcal{M'}}{\mathcal{M}}\mathcal{H} \right) \,  E^0  +
\frac {(E^{1q}-E^2_q)} {\sqrt{2{\cal M}}}\; \frac { {\rm tr} \left(4i (\gamma^i {\bf{\Psi}})_q  {\bb P}^i +
(\gamma^{ij} {\bf{\Psi}})_q  [{\bb X}^i, {\bb X}^j]\right) -  \frac{1}{2\mu^6} \frac{\mathcal{M'}}{\mathcal{M}}\mathfrak{H}\, i\nu_q  } { \left(1 + \frac{1}{\mu^6} \frac{\mathcal{M'}}{\mathcal{M}}\mathfrak{H} \right)}\,  . \qquad
\ee
with \be\label{frakH=} {\frak H}:= \text{tr}\left( {\bb P}^i {\bb P}^i \right)   + {1\over 16}
\text{tr}\left[ {\bb X}^i ,{\bb X}^j \right]^2 + 2\,  \text{tr}\left({\bb X}^i\, {\bf \Psi}\gamma^i {\bf{\Psi}}\right)\; .
\ee

Thus the final form of the equations for the matrix matter fields  is given by
\eqref{DXi==}, \eqref{DPi==} and \eqref{DPsi=} with $\text{d}\mathcal{K}$ substituted by the r.h.s. of \eqref{dcK==}.
This looks frighteningly complicated but, as we will show, the equations can be simplified essentially by fixing in a convenient manner the gauge symmetries of our dynamical system.
However, before turning to the gauge fixing, we have to discuss the mass of the mD$0$ system.

\subsection{ Mass of mD0 system and its center of mass motion}

It is instructive to calculate the canonical momentum for the center of energy coordinate function of mD$0$ system
\be\label{pmu=}
p_\mu = \left(m + \frac{2}{\mu^6} \frac{\mathcal{H}}{\mathcal{M}}\,\right) u_\mu^0=: \frak{M} u_\mu^0  \, .
\ee
The mass $ \frak{M}$ of the mD$0$ is defined by the square of this 10-momentum, $p_\mu p^\mu =  \mathfrak{M}^2$,  and thus is given by
\be\label{Mass=}
\mathfrak{M} = m + \frac{2}{\mu^6} \frac{\mathcal{H}}{\mathcal{M}}\,\, .
\ee
Note that it depends essentially on the choice of the positively definite function ${\mathcal{M}}(\mathcal{H}/\mu^6)$ in the action \eqref{SmM0=}.
However, as a consequence of  \eqref{dH=0}, this mass is a constant of motion,
\be\label{dfM=0}
\text{d}\frak{M}=0\; .
\ee

It is important that the mass depends on the relative motion of the mD$0$ constituents so that in our supersymmetric model the relative motion does influence the center of energy motion like it does in the purely bosonic dynamical system from \cite{Myers:1999ps}.

The constant parameter $m$ in the action \eqref{SmD0=} is thus the mass of the center of energy motion of mD$0$ the relative motion sector of which is in the ground state, i.e. with ${\cal H}=0$. As we will see below, the supersymmetric states of mD$0$ systems have this property.
Thus exciting the state of relative motion of the  mD$0$ constituents we inevitably increase the mass of the center of energy of mD$0$ system (remember  \eqref{cM>0}) and, as we will see below, inevitably break supersymmetry.

Now let us also observe  that, by definition of Cartan forms, $\text{d}u_\mu^0= u_\mu^i\Omega^i$ and hence Eq.\eqref{Omi=0} implies that on the mass shell
\be \label{du0=0}
\text{d}u_\mu^0=0\; .
\ee
This together with \eqref{dfM=0} implies that the momentum \eqref{pmu=} is a constant of motion
\be \text{d}p_\mu=0~. \ee

\subsection{Gauge fixing of the local $\text{SO}(9)$ and  $\text{SU}(N)$ symmetries}
\label{sec:convenientGF}

For the future use, let us also notice that on the surface of  Eq. \eqref{Omi=0} the derivatives of the orthogonal moving frame vectors
$u_\mu^i$ are decomposed on the set of these vectors only,
\be
\text{d}u_\mu^i= \Omega^{ij} u_\mu^j \; .
\ee
Furthermore, as $\Omega^{ij}$ transforms as the connection under local gauge \text{SO}$(9)$ symmetry of the mD$0$ action \eqref{SmD0=}, and as any 1-dimensional connection can be gauged away, we can always fix the gauge
\be\label{Omij=0}
\Omega^{ij}= 0 \ee
in which also 9 spacelike vectors of the moving frame become constant
\be\label{dui=0}
\text{d}u_\mu^i= 0  \; .
\ee
Similarly, we can use the $\text{SU}(N)$ gauge symmetry to fix the gauge in which the $\texttt{su}(N)$ valued gauge field vanishes,
\be
{\bb A}= 0\; ,
\ee
so that the covariant derivative $\text{D}=\text{d}\tau \text{D}_\tau$ reduces to time derivative,
\be
\text{D}{\bb X}^i= \text{d}\tau \frac{\text{d}} {\text{d}\tau} {\bb X}^i=  \text{d}\tau  \dot{{\bb X}}{}^i\; , \qquad  \text{D}{\bb P}^i= \text{d}\tau \frac{\text{d}}{\text{d}\tau} {\bb P}^i=  \text{d}\tau  \dot{{\bb P}}{}^i\; , \qquad
\text{D}{\mathbf{\Psi}}_q= \text{d}\tau \frac{\text{d}}{\text{d}\tau} {\mathbf{\Psi}}_q=  \text{d}\tau  \dot{\mathbf{\Psi}}_q\; .  \qquad
\ee

Notice that, as far as moving frame and spinor moving frame variables are now (proper-)time independent,
\be\label{E0=gauge}
E^0=\text{d}\tau E_\tau^0= \text{d}x^{{\bf 0}} -i \text{d}\theta^{1q}\,\theta^{1q}-i \text{d}\theta_q^{2} \, \theta_q^{2} \; , \qquad E^{q1}= \text{d}\theta^{1q}\; , \qquad E_q^{2}= \text{d}\theta_q^{2}\; , \qquad
\ee
and
\be\label{Ei=gauge}
E^i=\text{d}\tau E_\tau^i= \text{d}x^{{\bf i}} -i \text{d}\theta^{1q}\gamma^i_{qp}\theta^{1p}+i \text{d}\theta_q^{2} \gamma^i_{qp} \theta_p^{2} \; , \qquad
\ee
where
\be\label{x0=xu0}
x^{{\bf 0}}=x^\mu u_\mu^0\; , \qquad x^{{\bf i}}=x^\mu u_\mu^i\; , \qquad
\theta^{1q}=\theta^{\alpha 1} v_\alpha{}^q \; , \qquad  \theta^2_{q}=\theta^2_\alpha v^\alpha_q \;   \qquad
\ee
describe supersymmetric generalization of the co-moving coordinate system  of the center of mass of mD$0$-brane. 

\subsection{Gauge fixing of $\kappa$-symmetry and reparametrization symmetry}

\label{secGFk}

The last two of  consequences  \eqref{E0=gauge} of Eqs. \eqref{Omi=0} and the gauge fixing conditions \eqref{Omij=0}
allow to reduce \eqref{E1+E2=0} to
\be\label{dth1=-dth2}
 \text{d}\theta^{1q}= - \text{d}\theta_q^{2}\; , \qquad
\ee
This equation is clearly invariant under spacetime supersymmetry and $\kappa$-symmetry  which in coordinate basis \eqref{x0=xu0} implies
\be
\delta x^{{\bf 0}}=i\text{d}\theta^{1q}\, (\epsilon^{1q}- \kappa^q/\sqrt{2})+i\theta_q^{2} \, ( \epsilon_q^{2}+ \kappa^q/\sqrt{2}) \; , \qquad
\delta  \theta^{1q} =\epsilon^{1q}+ \kappa^q/\sqrt{2} \; , \qquad \delta \theta_q^{2} =  \epsilon_q^{2}- \kappa^q/\sqrt{2}\;  \qquad
\ee
and
\be
\delta E^0= -2i ( \text{d}\theta^{1q} -\text{d} \theta_q^{2}) \kappa^q/\sqrt{2}\; .
\ee

Clearly, the worldline supersymmetry  can be used to set one of two  (but not both) fermionic coordinate functions equal to zero (as the $\kappa$-symmetry of single D$0$-brane can be). Let us choose
\be\label{th2=0}
\theta_q^2=0 \; .
\ee
This gauge is preserved by the combination of supersymmetry and worldline supersymmetry which obey
\be\label{e2=k}
\epsilon^2_q = \kappa^q/\sqrt{2}\; .
\ee
This relation also implies that the parametric function of the worldline supersymmetry becomes a fermionic constant spinor at this stage.

The gauge choice \eqref{th2=0}   simplifies Eq.  \eqref{dth1=-dth2} to
\be\label{dth1=0}
 \text{d}\theta^{1q}= 0\; , \qquad
\ee
which in its turn reduces $E^{0}$ form \eqref{E0=gauge} to $\text{d}x^{{\bf 0}}$,
\be
E^{0}=\text{d}x^{{\bf 0}}\, ,
\ee
and
 $E^{i}$ form \eqref{Ei=gauge} to $\text{d}x^{{\bf i}}$,
\be
E^{i}=\text{d}x^{{\bf i}}\, .
\ee
This is supersymmetric because now
\be
\delta x^{{\bf 0}}= i (\epsilon^{1q}+\epsilon^2_q)\theta^{1q}\; , \qquad \delta x^{{\bf i}}= i (\epsilon^{1q}+\epsilon^2_q)\gamma^i_{qp}\theta^{1p}\;
\ee
and the r.h.s.-s of these expressions are constants (due to \eqref{dth1=0}, $\text{d}v_\alpha^q=0$ and $\text{d}\epsilon^{\alpha 1,2}=0$).

In this gauge with respect to the $\kappa$-symmetry   we can fix the gauge with respect to the reparametrization symmetry by setting
\be\label{dt=dx0}
\text{d}\tau= \text{d}x^{{\bf 0}}=E^0 \qquad \Rightarrow \qquad E_\tau^0= \dot{x}^{{\bf 0}}=1\;
\ee
still preserving the supersymmetry. We however find convenient to do not do this and preserve explicit $\tau$--reparametrization symmetry at the next stages.

\subsection{Gauge fixed form of the field equations}

Thus the above gauge fixing and the field equations for  the center of energy variables imply that
\bea\label{du=0=dv} \text{d}u^{0}_\mu=0\; , \qquad \text{d}u^{i}_\mu=0\; , \qquad  \text{d}v_\alpha^q=0\; ,  \\ \label{Ei=0b}
 E^i =\text{d}x^{{\bf i}} = 0\; , \\
 E^{1q}=\text{d}\theta^{1q}=0\; , \qquad E^2_q=0\; ,
\eea
and $E^0=\text{d}\tau \dot{x}^{\bf 0}$.
With this in mind the equations for the matrix fields reduce to
\bea \label{DXi=gf}
\dot{\mathbb{X}}^i &=& -\frac{2}{\mathcal{M}} \left(1 - \frac{1}{\mu^6}\frac{\mathcal{M'}}{\mathcal{M}}\mathcal{H} \right)\dot{x}^{\bf 0} \mathbb{P}^i - \frac{1}{\mu^6}\frac{\mathcal{M'}}{\mathcal{M}} \mathbb{P}^i  \dot{\mathcal{K}}\; , \qquad
\\ \label{DPi=gf}
\dot{\mathbb{P}}^i &=&\left[ \frac{2}{\mathcal{M}}  \left(1- \frac{1}{\mu^6} \frac{\mathcal{M'}}{\mathcal{M}}\mathcal{H} \right)\dot{x}^{\bf 0} + \frac{1}{\mu^6} \frac{\mathcal{M'}}{\mathcal{M}}\dot{\mathcal{K}} \right] \left(\frac{1}{16} \left[[\mathbb{X}^i, \mathbb{X}^j],\mathbb{X}^j  \right] - \gamma^i_{pr} \left\lbrace\mathbf{\Psi}_p, \mathbf{\Psi}_r \right\rbrace  \right)\; , \qquad
 \\
\label{DPsi=gf} \dot{\mathbf{\Psi}}_q &=& - \frac{i}{2\mathcal{M}} \left( \left(1 - \frac{1}{\mu^6} \frac{\mathcal{M'}}{\mathcal{M}}\mathcal{H}\right)\dot{x}^{\bf 0} +  \frac{\mathcal{M'}}{2\mu^6} \dot{\mathcal{K}}    \right) [(\gamma^i \mathbf{\Psi})_q, \mathbb{X}^i]  \; ,
\eea
where
\be\label{dcK=gf}
\dot{\mathcal{K}}= - \frac 2 {{\cal M}}  \frac {\mathfrak{H} }{ \left(1 + \frac{1}{\mu^6} \frac{\mathcal{M'}}{\mathcal{M}}\mathfrak{H} \right)}\,  \left(1- \frac{1}{\mu^6} \frac{\mathcal{M'}}{\mathcal{M}}\mathcal{H} \right) \, \dot{x}^{\bf 0}  \qquad
\ee
with ${\frak H}$ defined in \eqref{frakH=}.

Substituting \eqref{dcK=gf} we find that the final form of the gauge fixed Eqs. \eqref{DXi=gf}--\eqref{DPsi=gf} is
\bea\label{DXi==gf}
\dot{\mathbb{X}}{}^i &=& -\frac{2}{\mathcal{M}} \, \frac {\left(1 - \frac{1}{\mu^6}\frac{\mathcal{M'}}{\mathcal{M}}\mathcal{H} \right)} { \left(1 + \frac{1}{\mu^6} \frac{\mathcal{M'}}{\mathcal{M}}\mathfrak{H} \right)} \dot{x}^{\bf{0}}\, \mathbb{P}^i \; , \qquad
\\ \label{DPi==gf}
\dot{ \mathbb{P}}{}^i &=& \frac{2}{\mathcal{M}}\frac {\left(1 - \frac{1}{\mu^6}\frac{\mathcal{M'}}{\mathcal{M}}\mathcal{H} \right)} { \left(1 + \frac{1}{\mu^6} \frac{\mathcal{M'}}{\mathcal{M}}\mathfrak{H} \right)} \dot{x}^{\bf{0}} \,\left(\frac{1}{16} \left[[\mathbb{X}^i, \mathbb{X}^j],\mathbb{X}^j  \right] - \gamma^i_{pr} \left\lbrace\mathbf{\Psi}_p, \mathbf{\Psi}_r \right\rbrace  \right) \; ,\qquad \\
\label{DPsi==gf} \dot{\mathbf{\Psi}}_q &=& - \frac{i}{2\mathcal{M}} \, \frac {\left(1 - \frac{1}{\mu^6}\frac{\mathcal{M'}}{\mathcal{M}}\mathcal{H} \right)} { \left(1 + \frac{1}{\mu^6} \frac{\mathcal{M'}}{\mathcal{M}}\mathfrak{H} \right)}\, \dot{x}^{\bf 0}  [(\gamma^i \mathbf{\Psi})_q, \mathbb{X}^i]  \; .
\eea

Notice that, despite the possibility to fix the gauge $\dot{x}^0=1$ \eqref{dt=dx0}, we prefer to keep it unfixed to stress the invariance of our equations of motion under the $\tau$-reparametrizations, the fact which will be important for our discussion below.

\subsection{Relation of the relative motion equations of mD$0$ with  SYM equations }

Formally, we can define the new time variable by
\be\label{dt=}
\text{d}t = \text{d}x^{{\bf 0}} \,  \frac{2}{\mathcal{M}}\frac {\left(1 - \frac{1}{\mu^6}\frac{\mathcal{M'}}{\mathcal{M}}\mathcal{H} \right)} { \left(1 + \frac{1}{\mu^6} \frac{\mathcal{M'}}{\mathcal{M}}\mathfrak{H} \right)} \qquad \Leftrightarrow \qquad
\frac {\text{d}t(\tau)} {\text{d}\tau} = \dot{x}^{{\bf 0}} \,  \frac{2}{\mathcal{M}}\frac {\left(1 - \frac{1}{\mu^6}\frac{\mathcal{M'}}{\mathcal{M}}\mathcal{H} \right)} { \left(1 + \frac{1}{\mu^6} \frac{\mathcal{M'}}{\mathcal{M}}\mathfrak{H} \right)} \; ,
\ee
and write the above equations in terms of derivative $\frac{\text{d}}{\text{d}t}$ arriving at
\bea\label{DXi=SYM}
\frac{\text{d}}{\text{d}t} {\mathbb{X}}{}^i &=& -\, \mathbb{P}^i \; , \qquad
\\ \label{DPi=SYM}
\frac{\text{d}}{\text{d}t} { \mathbb{P}}{}^i &=& \frac{1}{16} \left[[\mathbb{X}^i, \mathbb{X}^j],\mathbb{X}^j  \right] - \gamma^i_{pr} \left\lbrace \mathbf{\Psi}_p, \mathbf{\Psi}_r \right\rbrace \; ,\qquad \\
\label{DPsi=SYM} \frac{\text{d}}{\text{d}t} \mathbf{\Psi}_q &=& - \frac{i}{4} \,   [(\gamma^i \mathbf{\Psi})_q, \mathbb{X}^i]  \;
\eea
which has the form of the equations of motion of $\text{d}=1$ reduction of 10D YM theory in the gauge ${\bb A}_0=0$.

Notice, however, that this procedure cannot be considered as reparametrization (1d general coordinate) transformations of the proper time $\tau$.
The simplest way to see this is to appreciate that the defining equation \eqref{dt=}  for the ``new time'' is actually invariant under the $\tau$-reparametrizations and hence cannot be a gauge fixing condition for this symmetry  \footnote{Actually the equations \eqref{DXi=gf}--\eqref{DPsi=gf} are manifestly reparametrization invariant themselves and thus cannot be transformed into
the other form \eqref{DXi=SYM}--\eqref{DPsi=SYM}
by $\tau$-reparametrizations.}.

The other, more physical observation is that a change of the form of the positively definite function ${\cal M}({\cal H})$ implies the changes of the physical characteristic of the system. Namely  such a passage will change the mass ${\frak M}$ of the mD$0$ system \eqref{Mass=} which is different for different forms of ${\cal M}({\cal H})$. Hence, such a change cannot be achieved by transformation of a gauge symmetry of the system.

Thus what we have found is not a kind of gauge equivalence, but  an interesting correspondence of the gauge fixed form of the relative motion equations of our mD$0$ system with the equation of maximally supersymmetric ${\cal N}=16$  $\text{d}=1$ SU($N$) SYM theory. Notice that the formal  definition \eqref{dt=} of new time variable involves the matrix fields ${\bb X}^i$, ${\bb P}^i$, ${\bf \Psi}_q$ and thus generically it is different for different solutions of the field equations. Indeed, although  ${\cal H}$, and hence ${\cal M}({\cal H})$, are constants on the mass shell, this is not the case for ${\frak H}={\frak H}({\bb X}^i, {\bb P}^i, {\bf \Psi}_q)$.
However, in particular case of solutions with ${\cal H}=0$, also ${\frak H}=0$, $t$ is related with
$x^{\textbf{0}}$ by a constant re-scaling and the correspondence between such solutions of relative motion  mD$0$ and SU($N$) SYM equations is  one-to-one.

As we will see in a moment, this is the case for bosonic supersymmetric solutions of mD$0$ equations.

\section{Supersymmetric bosonic solutions}
\label{sec:SUSYbosSol}

As we have discussed in sec. \ref{secGFk}, on the mass shell one of two center of energy fermions can be gauged away by the worldline supersymmetry ($\kappa$-symmetry), \eqref{th2=0}, and then the remaining fermionic coordinate function becomes constant \eqref{dth1=0}. Setting this also to zero, thus arriving at purely bosonic center of mass configuration with
\be\label{th1=0}
\theta^{\alpha 1}=0 \; , \qquad \theta_\alpha^2=0 \; ,
\ee
we find that such a choice  is still preserved by a half of the target (super)space supersymmetry with two spinor parameters expressed in terms of one SO(9) spinor parameter  of the worldline supersymmetry and spinor frame variables (which are constant in the gauge under consideration, \eqref{du=0=dv}) by
\be\label{e1=k}
\epsilon^{\alpha 1} = -\kappa^qv_q^\alpha /\sqrt{2}\; , \qquad \epsilon_\alpha^2 = \kappa^qv_\alpha^q/\sqrt{2}\;.
\ee

Then, after fixing the reparametrization invariance by setting $x^{{\bf 0}}(\tau)=\tau$,  the solution of the bosonic center of energy equations can be written in the form
\be\label{x=x0+pt}
x^\mu (\tau)  = x^\mu_0 +\frac {p^\mu }{\frak{M}}\tau \; ,
\ee
where $p^\mu$ and $\frak{M}$ are given in \eqref{pmu=} and \eqref{Mass=}. These are constants due to \eqref{dH=0} which is  the Noether identity for the reparametrization symmetry. Let us stress that  any of these bosonic solutions of the equations of center of energy motion preserves 1/2 of the type IIA spacetime supersymmetry, which can be then broken or preserved by the relative motion of the mD$0$ constituents.

Originally the matrix fields which describe the relative motion of the mD$0$ constituents are inert under spacetime supersymmetry. However, after gauge fixing of the worldline supersymmetry ($\kappa$-symmetry) by \eqref{th2=0}, this is not more the case since the parameter of the worldline supersymmetry, which does act on the matrix fields, becomes identified with the parameter of the second spacetime supersymmetry, Eq. \eqref{e2=k}. Furthermore, setting to zero the remaining fermionic coordinate function, \eqref{th1=0}, identifies with that also the parameter of the first spacetime supersymmetry \eqref{e1=k}, so that the action of spacetime supersymmetry on the matrix fields becomes nontrivial but coincide with the action of the worldline supersymmetry which is now parametrized by a constant fermionic $\kappa^q$.

Setting to zero the fermionic matrix field
\be\label{Psi=0} {\bf \Psi}_q=0
\ee
we can still preserve the worldline supersymmetry (completely or partially), and hence also a part of the spacetime supersymmetry if
the following Killing spinor equation (see \eqref{susy=Psi} in Appendix A)
\be
\label{susy=Psi=b}\delta_\kappa {\mathbf{\Psi}}_q=-\frac 1 {2\sqrt{{\cal M}}}\, (\kappa\gamma^{i})_q {\bb P}^i  - \frac {i}{16\sqrt{{\cal M}}}\, (\kappa\gamma^{ij})_q [ {\bb X}^i, {\bb X}^j]=0
\;
\ee
has a nontrivial solution. Notice that the contribution of the nonvanishing function ${\cal M}({\cal H})$ can be factored out thus reducing the equation to
\be
\label{susy=Psi=bb}  \kappa^p {\bb L}_{pq}:= \kappa^p \left(\gamma^{i}_{pq} {\bb P}^i  - \frac {i}{8}\, (\gamma^{ij})_{pq} [ {\bb X}^i, {\bb X}^j]\right)=0\;
\;
\ee
which is formally the same as that defining the potentially preserved supersymmetry in the case of (the first order formulation of) the maximally supersymmetric SU($N$) SYM theory\footnote{The word formally refers to the fact that the relation between ${\mathbb P}^i$ and $\dot{{\mathbb X}}^i$ in the case of the generic mD$0$ action
includes the contribution of ${\cal M}({\cal H})$ and, if ${\cal M}^\prime \not= 0$, also of
${\mathbb P}^i$ and ${{\mathbb X}}^i$  inside of $\frak{H}$ function \eqref{frakH=}, see \eqref{DPi==gf}.}. It also formally coincides with the equation defining the supersymmetry which can be  preserved by purely bosonic solutions of 11D mM$0$ equations \cite{Bandos:2012jz} so that our discussion below will be close to the one in \cite{Bandos:2012jz}.

The preservation of the 1/2 of the target superspace supersymmetry implies, in the light of \eqref{e1=k}, that
Eq. \eqref{susy=Psi=bb} does not impose any restriction on the parameter $\kappa^q$ which implies that the matrix ${\bb L}_{pq}$ vanishes, $\gamma^{i}_{pq} {\bb P}^i  - \frac {i}{8}\, (\gamma^{ij})_{pq} [ {\bb X}^i, {\bb X}^j]=0$. This in its turn implies that
\be\label{vacuum}
 {\bb P}^i=0\; , \qquad [ {\bb X}^i, {\bb X}^j]=0 \, ,
\ee
which describes the vacuum of the relative motion of the mD$0$ constituents. In it
${\cal H} =0$ and, actually, \eqref{vacuum} can be obtained from this equation.

Now let us check whether other parts of spacetime supersymmetry can be preserved. To this end we have to study
Eq. \eqref{susy=Psi=bb} which is actually a  set of 16  traceless $N\times N$ matrix equations. It is convenient to
extract from this just 16  equations by multiplying it by the matrix   $\tilde{{\bb L}}_{qr}:= \left(\gamma^{i}_{qr} {\bb P}^i  + \frac {i}{8}\, (\gamma^{ij})_{qr} [ {\bb X}^i, {\bb X}^j]\right)$, which differs by relative sign from the matrix ${{\bb L}}_{qr}$ in  \eqref{susy=Psi=bb}, and tracing the result with respect to the $\text{SU}(N)$ indices. In such a way we arrive at
\be
\label{kpL2tL2=0}  \kappa^p {\rm tr} ({\bb L}\tilde{\bb L})_{pq}:=
\kappa^q \left({\rm tr}( {\bb P}^i{\bb P}^i)  - \frac {1}{32}\, {\rm tr}([ {\bb X}^i, {\bb X}^j]^2)\right) + \frac {i}{4}\, \kappa^p  (\gamma^{j})_{pq}{\rm tr}( {\bb P}^i [ {\bb X}^i, {\bb X}^j])=0\; ,
\;
\ee
where we have used $\gamma^{ijkl} {\rm tr}(\left[\mathbb{X}^i,\mathbb{X}^j]\left[\mathbb{X}^k, \mathbb{X}^l \right] \right)=\gamma^{ijkl}~ {\rm tr}\left( \mathbb{X}^i \left[\mathbb{X}^j, \left[\mathbb{X}^k, \mathbb{X}^l \right] \right] \right)=0$ (which follows from the Jacobi identity $\left[\mathbb{X}^{[j}, \left[\mathbb{X}^k, \mathbb{X}^{l]} \right] \right]\equiv 0$). Now one can recognize in the multiplier of the first term the  bosonic limit of the relative motion Hamiltonian   $\mathcal{H}$  \eqref{cH=} multiplied by 2 and also appreciate that the coefficient for the second term vanishes as a consequence of the (bosonic limit of the) Gauss constraint \eqref{Gauss=}, ${\rm tr}( {\bb P}^i [ {\bb X}^i, {\bb X}^j])={\rm tr}( [{\bb P}^i , {\bb X}^i] {\bb X}^j)=0$. Thus, if we use this constraint, the final form of the consistency condition \eqref{kpL2tL2=0}  for the supersymmetry preservation by mD$0$ configuration simplifies to
\begin{equation}\label{kcH=0}
\kappa^q \mathcal{H} = 0~. \qquad
\end{equation}
Hence the BPS equations determining the supersymmetric pure bosonic solution of the mD$0$ equations is just the vanishing of the (bosonic limit of the) relative motion Hamiltonian,
\begin{eqnarray}
\label{cHb=0} {\cal H} &=& {1\over 2} \text{tr}\left( {\bb P}^i {\bb P}^i \right)   - {1\over 64}
\text{tr}\left[ {\bb X}^i ,{\bb X}^j \right]^2  =0\; .  \quad
  \end{eqnarray}
As we have already noticed, the general solution of this equation is given by the vacuum \eqref{vacuum} so that only 1/2 of the spacetime supersymmetry can be conserved by the mD$0$ system.

Furthermore, this allows to conclude that any supersymmetric solution of the maximally supersymmetric SU($N$) SYM theory gives rise to the set of supersymmetric solutions of the mD$0$ equations with arbitrary nonvanishing
function ${\cal M}({\cal H})$ since the relative motion mD$0$ equations then differ from the SYM equations by rescaling of the time variable by the constant factor ${\cal M}(0)/2$. Here we are speaking about a family of solution since these can have different characteristics of the center of mass motion $x^\mu_0$ and $p^\mu = \frak{M}u^{0\mu}$ in \eqref{x=x0+pt}. As far as the supersymmetric solutions of relative motion equations of mD$0$ constituents is concerned, these are in one-to-one correspondence with such solutions of the SU($N$) SYM equations.

The correspondence of the relative motion equations of the mD$0$ brane constituents and the maximally supersymmetric $\text{d}=1$ ${\cal N}=16$ SU($N$) SYM equations can be also used to search for non-supersymmetric solutions of the former. See Appendix D for an example.

\section{Conclusion}
\label{sec:conclusions}

In this paper we have studied the properties of the dynamical system described by recently constructed in \cite{Bandos:2022uoz} doubly supersymmetric nonlinear action \eqref{SmD0=} which possesses the properties expected from the action for multiple super-D$0$-brane (mD$0$) system (hence the name candidate mD$0$ action or simply mD$0$ action which we use above).

Doubly supersymmetric means that the action possesses both spacetime (or more precisely, target superspace) supersymmetry and worldline supersymmetry generalizing the famous $\kappa$-symmetry of the action for single D$0$-brane\footnote{This was found for the first time in \cite{deAzcarraga:1982dhu} for ${\cal N}=2$ $\text{D}=4$ massive superparticle, lower dimensional counterpart of 10D D$0$-brane \cite{Bergshoeff:1996tu}.}.
This property guarantees that the ground state of the dynamical system preserves a part (one half) of the target space supersymmetry, and thus is a stable 1/2 BPS state, the property expected for mD$0$ brane system.

The action of \cite{Bandos:2022uoz}, Eq. \eqref{SmD0=}, which is under study in this paper, includes arbitrary positive definite function ${\cal M}({\cal H}/\mu^6)$ of the relative motion Hamiltonian. The previously proposed in \cite{Bandos:2018ntt} simpler candidate mD$0$ action can be obtained by choosing  ${\cal M}$ to be equal to the constant mass parameter $m$ which appears in the center of energy part of the action, ${\cal M}({\cal H}/\mu^6)=m$. This center of energy part of the action coincides formally with the spinor moving frame action \cite{Bandos:2000tg} of single D$0$-brane with mass $m$.

As we have shown in this paper, the action \eqref{SmD0=} with other particular  choice,
${\cal M} =\frac m 2+  \sqrt{\frac {m^2} 4+\frac {{\cal H}}{\mu^6}}$ (where $\mu$ is a parameter of dimension of mass entering the relative motion sector of the action and characterizing its interaction with the center of energy sector)
can be  obtained by dimensional reduction from the 11D mM$0$ action of \cite{Bandos:2012jz}, Eq. \eqref{SmM0=} (this fact was announced in \cite{Bandos:2022uoz}).

We have also obtained the complete set of equations of motion which follows from the action \eqref{SmD0=}
with an arbitrary positive ${\cal M}({\cal H}/\mu^6)$
and show that their part describing the center of energy motion formally coincides with the set of  equations for single D$0$-brane but with  effective mass
$\mathfrak{M} = m + \frac{2}{\mu^6} \frac{\mathcal{H}}{\mathcal{M}}$ \eqref{Mass=}
expressed in terms of the above mentioned positive definite function  ${\cal M}({\cal H}/\mu^6)$.
The complete gauge invariant form of the relative motion equations is quite  complicated but we have shown that they imply that the relative motion Hamiltonian is a constant of motion, d${\cal H}=0$. This relation, which is the Noether identity manifesting the reparametrization symmetry (1d general coordinate invariance) of our mD$0$ action \eqref{SmD0=}, is particularly important as it guarantees that the effective mass of mD$0$ system is constant.

The relative motion equations simplify essentially in a certain gauge fixed on the center of energy variables  and 1d SU($N$) gauge field. In this gauge we have established an interesting relation of the gauge fixed form of our equations of relative motion with the equations of ${\cal N}=16$ $\text{d}=1$ SU($N$) SYM model.
This relation, which is valid for arbitrary positive ${\cal M}({\cal H}/\mu^6)$, does not imply a gauge equivalence. This fact can be seen from that it can be used to relate the (equations of relative motion of the) models with different  ${\cal M}$ and hence with different values of the mass $\frak{M}$ of our mD$0$ system, Eq. \eqref{Mass=}.

The established relation allows to show that all supersymmetric bosonic solutions of our mD$0$ equations (in its relative motion sector) are given essentially by SUSY solutions of the SYM model.
The BPS equations which are obeyed by that supersymmetric solutions are shown to reduce to single equation of  vanishing of the relative motion Hamiltonian ${\cal H}=0$. Hence, according to \eqref{Mass=}, the effective mass  $\frak{M}$ of the configurations described by these supersymmetric solutions
are given by center of energy parameter $m$.

Thus, in other words, for any choice of positive function ${\cal M}({\cal H}/\mu^6)$
the spectrum of BPS sector of our mD$0$ model  essentially coincides (as far as the relative motion sector is considered) with the set of vacua of maximally supersymmetric ${\cal N}=16$ $\text{d}=1$ SU($N$) SYM theory. The established correspondence allows also to study the non-supersymmetric solutions of our mD$0$ equations using the knowledge of the SYM solutions (see Appendix D for a simple example).

The enigma of surprising multiplicity of massive $p=0$ supersymmetric objects still cannot be claimed as resolved \footnote{Notice that the set of these includes, besides our mD$0$-models  with different positive ${\cal M}({\cal H}/\mu^6)$, also beautiful $0$-brane model of \cite{Panda:2003dj} including arbitrary function of relative motion variables
$\bar{{\cal M}}({\mathbb X}^i,\mathbf{\Psi}_q)$.}.
However,  our present  study of the equations following from the candidate mD$0$ actions \eqref{SmD0=} \cite{Bandos:2022uoz} has allowed to establish an interesting relation  of the gauge fixed equations
for any ${\cal M}({\cal H}/\mu^6)$ and maximally supersymmetric SYM equations which, in its turn, suggests a possible reason why such a multiplicity becomes possible. In particular, it allows to show that BPS spectra of the model \eqref{SmD0=}  with different ${\cal M}({\cal H}/\mu^6)$ coincides and in its relative motion sector is given by the BPS spectrum of the SYM model.

One of the most interesting directions of the  development of our approach is to attack the problem of its generalization for the case of higher $p$ multiple D$p$-brane system, beginning from $p=1$ case of multiple Dirichlet strings (mD$1$). This simplest case is particularly interesting for an attempt to indicate what animal in the
10D type IIA $0$-brane Zoo does describe the true mD$0$ system as this has to be related to mD$1$ by T-duality transformations. Such a study could also deepen our comprehension in T-duality.

In this respect it is also interesting to note the appearance of an indication of a new possible inhabitant of  the type II $1$-brane Zoo. It comes from \cite{Brennan:2019azg} where a DBI--like non-Abelian 2d model  was obtained recently  by the so-called $T\bar{T}$- deformation of 2d non-Abelian SYM model.  It was noticed in \cite{Brennan:2019azg} that its properties are quite different from non-Abelian generalization of DBI action which would describe mD1 system, but its origin in $T\bar{T}$- deformation suggests, using the arguments from \cite{Baggio:2018rpv,Chang:2018dge,Chang:2019kiu,Coleman:2019dvf}, en existence of possible maximally supersymmetric generalizations. An interesting direction  for future study is to construct explicitly such supersymmetric generalization, to try to extend it to an action for new 10D multiple $1$-brane system and to find what a representative of supersymmetric multiple  $0$-brane
family is related to it by T-duality.

A search for generalization of our approach to mD1 case, as well as the above mentioned search for new exotic non-Abelian $1$-branes,  will require the use of the appropriate spinor moving frame (Lorentz harmonic) formalism which was developed
in \cite{Bandos:1991bh,Bandos:1992np,Bandos:1992ze}. This is suggested by a special role which is played by
$\text{SO}(1,9)/\text{SO}(9)$ spinor moving frame formalism in  construction   of our candidate mD$0$ actions and in studying their properties
\footnote{To stress the importance of the spinor moving frame variables, let us discuss the way of possible derivation of our model \eqref{SmD0=} starting from  maximally supersymmetric $\text{d}=1$  SU($N$) SYM model  which possesses $\text{SO}(16)$ symmetry and rigid ${\cal N}=16$ supersymmetry,
or from its nonlinear deformations, preserving all the supersymmetry and at least $\text{SO}(9)$ part of $\text{SO}(16)$.
First we observe that  spinor moving frame variables allow to convert the SO(9) spinor index of the fermionic parameter of rigid
SYM supersymmetry into the SO(1,9) Majorana-Weyl  spinor ($\text{Spin}(1,9)$) index which is carried by coordinate of target superspace and also by a parameter of $\kappa$-symmetry of single D$0$-brane. We can also find that
they can be used to provide us with a composite supergravity induced by embedding of the D$0$-brane worldline into the target superspace. Then we add the single super-D$0$-brane action to the SYM action coupled to supergravity, and thus possessing reparametrization symmetry and local supersymmetry due to this coupling, and arrive at our candidate mD$0$ action. For the simplest case of ${\cal M}=m$ this program was realized in \cite{Bandos:2018ntt}.}.

\acknowledgments{The work has been partially supported by Spanish AEI MCIN and FEDER (ERDF EU) under  grant PID2021-125700NB-C21  and by the Basque Government Grant IT1628-22. We are thankful to Dmitri Sorokin for the interest to this work and useful discussions.
 }

\appendix

\section{Worldline supersymmetry transformations  }

The action of the worldline gauge supersymmetry on the center of energy variables imitates the action of the $\kappa$-symmetry on the variables of single D$0$-brane
\bea\label{kappa=}
&& \delta_\kappa \theta^{1 \alpha}=  {\kappa^q} v_q^\alpha/{\sqrt{2}} \; , \qquad \delta_\kappa \theta_{\alpha}^{2}= -  {\kappa^q} v_\alpha{}^q/{\sqrt{2}} \; , \qquad \nonumber   \\ && \delta_\kappa x^\mu =i\delta_\kappa\theta^1 \sigma^\mu \theta^1+i\delta_\kappa\theta^2 \tilde{\sigma}{}^\mu \theta^2  \; , \qquad  \nonumber   \\
 && \delta_\kappa v_\alpha^q=0\qquad \Rightarrow \qquad  \delta_\kappa u_{{\mu}}^{0} =0= \delta_\kappa u_{{\mu}}^{i}\; .
 \eea
The worldline supersymmetry transformations of the matrix matter fields are \cite{Bandos:2022uoz}
 \bea\label{susy=X}
&& \delta_\kappa {\bb X}^i  = \frac {4i}{\sqrt{{\cal M}}}\, \kappa\gamma^i{\mathbf{\Psi}}   + \frac 1 {\mu^6} \,  \frac {{\cal M}^\prime}{{\cal M}} \, \delta_\kappa {\cal H}\;   {\bb X}^i -
\frac 1 {\mu^6} \,  \frac {{\cal M}^\prime}{{\cal M}} \, \Delta_\kappa {\cal K}\,  {\bb P}^i \; , \qquad \\
\label{susy=P} && \delta_\kappa {\bb P}^i =  - \frac {1}{\sqrt{{\cal M}}}\, [\kappa\gamma^{ij}{\mathbf{\Psi}},  {\bb X}^j]
  -\frac 1 {\mu^6} \,  \frac {{\cal M}^\prime}{{\cal M}} \, \delta_\kappa {\cal H}  {\bb P}^i
  +\frac 1 {\mu^6} \,  \frac {{\cal M}^\prime}{{\cal M}} \Delta_\kappa {\cal K}
 \left( \frac 1 {16} [[ {\bb X}^i, {\bb X}^j], {\bb X}^j]-\gamma^i_{pq} \{{\mathbf{\Psi}}_p, {\mathbf{\Psi}}_q\} \right) ,\;  \\
\label{susy=Psi}&& \delta_\kappa {\mathbf{\Psi}}_q=-\frac 1 {2\sqrt{{\cal M}}}\, (\kappa\gamma^{i})_q {\bb P}^i  - \frac {i}{16\sqrt{{\cal M}}}\, (\kappa\gamma^{ij})_q [ {\bb X}^i, {\bb X}^j]
 - \frac {i}  {4\mu^6} \,  \frac {{\cal M}^\prime}{{\cal M}} \, \Delta_\kappa {\cal K}\, [(\gamma^{i}{\mathbf{\Psi}})_q , {\bb X}^i]\;
\eea
where
\bea \label{kappaH=}
\delta_\kappa {\cal H} =  \frac 1{ \sqrt{{\cal M}}}\, \frac {{\rm tr} \left(\kappa^q{\mathbf{\Psi}}_q\left( [{\bb X}^i, {\bb P}^i] -4i\{{\mathbf{\Psi}}_{q}, {\mathbf{\Psi}}_{q}\}\right) \right) } {1+\frac 1 {\mu^6} \,  \frac {{\cal M}^\prime}{{\cal M}} \, {\frak H} } \; \qquad
\eea
with
\be\label{frakH=2} {\frak H}:= \text{tr}\left( {\bb P}^i {\bb P}^i \right)   + {1\over 16}
\text{tr}\left[ {\bb X}^i ,{\bb X}^j \right]^2 + 2\,  \text{tr}\left({\bb X}^i\, {\mathbf{\Psi}}\gamma^i {\bf{\Psi}}\right)\;
\ee
is the worldline supersymmetry variation of the relative motion Hamiltonian \eqref{cH=} and
\bea \label{kappaK=}
\Delta_\kappa {\cal K} = \frac 1{ 2\sqrt{{\cal M}}}\, \frac { {\rm tr}  \left(4i (\kappa\gamma^i {\bf{\Psi}}) {\bb P}^i + {5\over 2}
(\kappa\gamma^{ij} {\bf{\Psi}})  [{\bb X}^i, {\bb X}^j]  \right)}{1+\frac 1 {\mu^6} \,  \frac {{\cal M}^\prime}{{\cal M}} \, {\frak H} } . \;
\eea
Finally, the transformations of the worldline gauge field are
\bea\label{susy=A}
\delta_\kappa {\bb A} &=&  - \frac 2 {{\cal M}\sqrt{{\cal M}}}\, E^0\,  (\kappa^q{\mathbf{\Psi}}_q) \frac {\left(1- \frac 1 {\mu^6} \, \frac {{\cal M}^\prime}{{\cal M}}{\cal H}\right)}{\left(1+ \frac 1 {\mu^6} \, \frac {{\cal M}^\prime}{{\cal M}}\, {\frak H}\right)} + \frac 1 {\sqrt{2}{\cal M}} \, (E^{1q}-E_q^2)(\gamma^i\kappa)_q \,{\bb X}^i - \qquad  \nonumber \\
& -&  (E^{1q}-E_q^2)\,  \frac 1 {\mu^6} \, \frac {{\cal M}^\prime}{\sqrt{2}{\cal M}^2} \,  \frac  1 {\left(1+ \frac 1 {\mu^6} \, \frac {{\cal M}^\prime}{{\cal M}}\, {\frak H}\right)}  \kappa^p\, {{\mathbf{\Psi}}_{(q}\,{\rm tr}\left(4i (\gamma^i {\mathbf{\Psi}})_{p)} {\bb P}^i +\frac 5 2
(\gamma^{ij} {\bf{\Psi}})_{p)}  [\mathbb{X}^i, {\bb X}^j]  \right) }\,  .
\eea

\section{Complete set of gauge invariant  mD0 equations }

Center of mass equations are the same as in the case of single D$0$-brane,
\begin{eqnarray}\label{Ei=0A}
 E^i  = 0\; ,
\\ \label{E1+E2=0A}
E^{1q} + E^2_{q}  =0  \; ,
\\ \label{Omi=0A}
\Omega^i = 0 \, .
\end{eqnarray}

The relative motion equations are the Gauss constraint
\begin{equation}\label{Gauss=A}
[\mathbb{X}^i, \mathbb{P}^i] = 4i \left\lbrace {\mathbf{\Psi}}_q, {\mathbf{\Psi}}_q \right\rbrace~.
\end{equation}
and
\bea\label{DXi==A}
\text{D}\mathbb{X}^i &=& -\frac{2}{\mathcal{M}} \frac{\left(1-\frac{1}{\mu^6} \frac{\mathcal{M}'}{\mathcal{M}} \mathcal{H} \right)}{\left(1+\frac{1}{\mu^6} \frac{\mathcal{M}'}{\mathcal{M}} \mathfrak{H} \right)} E^0 \mathbb{P}^i + \frac {(E^{1q}-E^2_q)} {\sqrt{2{\cal M}}} \left[4i(\gamma^i \mathbf{\Psi})_q -  \frac{\frac{1}{2 \mu^6} \frac{\mathcal{M}'}{\mathcal{M}} \text{tr}\left(4i (\gamma^j \mathbf{\Psi})_q \mathbb{P}^j + \frac{5}{2}(\gamma^{jk} \mathbf{\Psi})_q [\mathbb{X}^j,\mathbb{X}^k] \right)}{1+\frac{1}{\mu^6} \frac{\mathcal{M}'}{\mathcal{M}} \mathfrak{H}}\mathbb{P}^i \right]\\
\label{DPi==A} \text{D} \mathbb{P}^i &=& \frac{2}{\mathcal{M}} \frac{\left(1-\frac{1}{\mu^6} \frac{\mathcal{M}'}{\mathcal{M}} \mathcal{H} \right)}{\left(1+\frac{1}{\mu^6} \frac{\mathcal{M}'}{\mathcal{M}} \mathfrak{H} \right)}E^0 \left(\frac{1}{16} \left[[\mathbb{X}^i, \mathbb{X}^j],\mathbb{X}^j  \right] - \gamma^i_{pr} \left\lbrace \mathbf{\Psi}_p, \mathbf{\Psi}_r \right\rbrace  \right) - \frac{\left(E^{1q}- E^2_{q}\right)}{\sqrt{2 \mathcal{M}}} [\left(\gamma^{ij} \mathbf{\Psi} \right)_q, \mathbb{X}^j]
\nonumber \\
&&+ \frac{\left(E^{1q}- E^2_{q}\right)}{\sqrt{2 \mathcal{M}}}\,\frac{1}{2 \mu^6} \frac{\mathcal{M}'}{\mathcal{M}}\,  \frac{\text{tr}\left(4i (\gamma^k \mathbf{\Psi})_q \mathbb{P}^k + \frac{5}{2}(\gamma^{kl} \mathbf{\Psi})_q [\mathbb{X}^k,\mathbb{X}^l] \right)}{1+\frac{1}{\mu^6} \frac{\mathcal{M}'}{\mathcal{M}} \mathfrak{H}} \left(\frac{1}{16} \left[[\mathbb{X}^i, \mathbb{X}^j],\mathbb{X}^j  \right] - \gamma^i_{pr} \left\lbrace \mathbf{\Psi}_p, \mathbf{\Psi}_r \right\rbrace  \right),\\
\label{DPsi==A} \text{D}\mathbf{\Psi}_q &=& - \frac{i}{2\mathcal{M}} \frac{\left(1-\frac{1}{\mu^6} \frac{\mathcal{M}'}{\mathcal{M}} \mathcal{H} \right)}{\left(1+\frac{1}{\mu^6} \frac{\mathcal{M}'}{\mathcal{M}} \mathfrak{H} \right)}E^0 [(\gamma^i \mathbf{\Psi})_q, \mathbb{X}^i] - \frac{\left(E^{1q}- E^2_{q}\right)}{2\sqrt{2 \mathcal{M}}} \left(\gamma^i_{pq} \mathbb{P}^i + \frac{i}{8} \gamma^{ij}_{pq} [\mathbb{X}^i,\mathbb{X}^j] \right)  \nonumber  \\ && - \frac{\left(E^{1p}- E^2_{p}\right)}{\sqrt{2 \mathcal{M}}} \, \frac{i}{8 \mu^6} \frac{\mathcal{M}'}{\mathcal{M}}\, \frac{\text{tr}\left(4i (\gamma^j \mathbf{\Psi})_p \mathbb{P}^j + \frac{5}{2}(\gamma^{jk} \mathbf{\Psi})_p [\mathbb{X}^j,\mathbb{X}^k] \right)}{1+\frac{1}{\mu^6} \frac{\mathcal{M}'}{\mathcal{M}} \mathfrak{H}} [(\gamma^i \mathbf{\Psi})_q, \mathbb{X}^i] \; .
\eea

Writing the relative motion equations we have already taken into account the fact that they and the center of mass equations  imply the preservation of the relative motion Hamiltonian \eqref{cH=},
\be\label{dH=0A}
\text{d}{\cal H}=0\; .
\ee
This is the Noether identity for the reparametrization symmetry. The Noether identity for the worldline supersymmetry,
\be\label{Dnu=EHA}
i\text{D}\nu_q= \frac {2\sqrt{2}}{\sqrt{{\cal M}}}(E^{1q}-E_q^2){\cal H}\;
\ee
where $i \nu_q$ is defined in \eqref{inu=},  also holds.

\section{Convenient representations for 11D gamma-matrices }
\renewcommand{\theequation}{C.\arabic{equation}}
\setcounter{equation}0

\label{AppGamma}

\subsection{$\text{SO}(1,9)$ covariant representation}

\be\label{G11=s10}
\Gamma^\mu_{\underline{\alpha}\underline{\beta}} = \left(\begin{matrix} \sigma^{\mu}_{\alpha\beta} & 0\cr
 0 & \tilde{\sigma}{}^{\mu\alpha\beta}
\end{matrix}\right)\; , \qquad \Gamma^*_{\underline{\alpha}\underline{\beta}} = \left(\begin{matrix} 0 & - \delta_{\alpha}{}^{\beta} \cr
- \delta_{\beta}{}^{\alpha}  & 0
\end{matrix}\right)\; , \qquad
\ee
\be\label{tG11=s10}
\tilde{\Gamma}^{\mu \underline{\alpha}\underline{\beta}} = \left(\begin{matrix}  \tilde{\sigma}{}^{\mu\alpha\beta}& 0\cr
 0 & \sigma^{\mu}_{\alpha\beta}
\end{matrix}\right)\; , \qquad \tilde{\Gamma}^{*\underline{\alpha}\underline{\beta}} = \left(\begin{matrix} 0 & \delta_{\beta}{}^{\alpha} \cr
\delta_{\alpha}{}^{\beta} & 0
\end{matrix}\right)\; , \qquad
\ee
\be\label{C11=10}
C_{\underline{\alpha}\underline{\beta}} = i\,\left(\begin{matrix} 0 &\delta_{\alpha}{}^{\beta} \cr
 -\delta_{\beta}{}^{\alpha}  & 0
\end{matrix}\right)\; , \qquad C^{\underline{\alpha}\underline{\beta}} = i\,\left(\begin{matrix} 0 & \delta_{\beta}{}^{\alpha} \cr
-\delta_{\alpha}{}^{\beta} & 0
\end{matrix}\right)\; , \qquad
\ee
\be
\tilde{\Gamma}^{\underline{\mu}}=C {\Gamma}^{\underline{\mu}}C\; .
\ee

\subsection{$\text{SO}(1,1)\times \text{SO}(9)$ covariant representation}

\begin{eqnarray}
 \label{11DGC=} & (\Gamma ^{\#})_{\underline{\alpha}\underline{\beta}}
 = \left( \begin{matrix}2\delta_{pq} & 0
 \cr 0 & 0 \end{matrix}  \right) =  \tilde{\Gamma}{}^{=\underline{\alpha}\underline{\beta}} \;  , \qquad  (\Gamma ^{=})_{\underline{\alpha}\underline{\beta}}
 = \left( \begin{matrix} 0 & 0
 \cr 0 & 2\delta_{pq}  \end{matrix}  \right) =  \tilde{\Gamma}{}^{\# \underline{\alpha}\underline{\beta}}  \;  , \qquad \\
  & (\Gamma ^{i})_{\underline{\alpha}\underline{\beta}}
 = \left( \begin{matrix} 0 & \gamma^{i}_{pq}
 \cr \gamma^{i}_{pq} & 0 \end{matrix}  \right)=-  \tilde{\Gamma}{}^{i \underline{\alpha}\underline{\beta}} \;  , \qquad \\
\label{11DC=} & C_{\underline{\alpha}\underline{\beta}}  =- C_{\underline{\beta}\underline{\alpha}}
= \left( \begin{matrix} 0 & i\delta_{pq}
 \cr -i\delta_{pq} & 0
\cr\end{matrix} \right)=  (C^{-1}){}^{\underline{\alpha}\underline{\beta}}=:  C^{\underline{\alpha}\underline{\beta}}\;  . \qquad
\end{eqnarray}

\section{A non-supersymmetric solution of mD$0$ equations}
\renewcommand{\theequation}{D.\arabic{equation}}
\setcounter{equation}0

 As an example of using the correspondence with 1d SYM of the relative motion equations of mD$0$ system to find non-supersymmetric  solutions of these,  let us discuss the SYM solution  used in \cite{Brahma:2022dsd} to study the possibility to describe cosmology in the frame of BFSS matrix model \cite{Banks:1996vh}. It uses the ansatz
\be \label{Xi=aY}
{\bb X}^i (\tau) = a(\tau)  {\bb Y}^i \,
\ee
where ${\bb Y}^i$ are nine constant traceless $N\times N$ matrices which obey
\be
[[{\bb Y}^i, {\bb Y}^j]{\bb Y}^j] =16\lambda {\bb Y}^i
\ee
with some constant $\lambda$.

Let us search for a solution of our equations of motion with this ansatz.
Eq. \eqref{DXi==gf}, after fixing the gauge $\dot{x}^{{\bf 0}}=1$, implies that
\be\label{Pi=bY}
 {\bb P}^i (\tau) = b(\tau)  {\bb Y}^i
\ee
where $b(\tau) $ should be found from solving Eq. \eqref{DXi==gf} with \eqref{Xi=aY} and \eqref{Pi=bY}.
Then, with  the above ansatz
\be\label{H=anstaz}
{\cal H}=c\left(b^2+\frac \lambda {2} a^4\right) \; , \qquad {\frak H}=2{\cal H} +\frac 3 {32}{\rm tr} ([{\bb X}^i , {\bb X}^i ]^2 )=2 {\cal H} - 3c\lambda a^4 = 2c\left(b^2-\lambda  a^4\right) \; , \qquad
\ee
where
\be
c=\frac 1 2 {\rm tr}({\bb Y}^i{\bb Y}^i)\; ,
\ee
 so that the straightforward approach to  \eqref{DXi==gf} does not look too promising.
However, at this stage we can use the equation \eqref{dH=0} stating that on the mass shell
${\cal H}$ is constant and conclude from \eqref{H=anstaz} that
\be\label{b2=}
 b^2= \frac {{\cal H}} c- \frac \lambda {2} a^4 \; , \qquad {\frak H}=2{\cal H} -\frac {3\lambda} {2} a^4 \; . \qquad
\ee

Furthermore, the approach in \cite{Brahma:2022dsd} allows for  the field-dependent redefinition of the time variable similar to \eqref{dt=} so that we use this with the chosen ansatz to define
\be\label{dt=2}
\text{d}t = \text{d}\tau  \,  \frac{2}{\mathcal{M}}\frac {\left(1 - \frac{1}{\mu^6}\frac{\mathcal{M'}}{\mathcal{M}}\mathcal{H} \right)} { \left(1 + \frac{1}{\mu^6} \frac{\mathcal{M'}}{\mathcal{M}}\left(2{\cal H} -\frac {3\lambda} {2} a^4(\tau)\right)  \right)} \qquad
\ee
and to obtain in such a way  the equations
\be
\frac{\text{d}}{\text{d}t}a=- b\; , \qquad \frac{\text{d}^2 a}{\text{d}t^2}=-\frac \lambda  a^3\; . \qquad
\ee
Following \cite{Brahma:2022dsd}, we can now multiply the last  equation by
$ \frac {d a} {dt}$ and integrate it over $dt$ arriving at
\be
\frac 1 2 \left(\frac{\text{d}a}{\text{d}t}\right)^2= - \frac \lambda 4 a^4 +k
\ee
with constant $k$. Next we can redefine the time variable once more
\be\label{dtt}
\text{d}\tilde{t}= \text{d}t a(t)
\ee
thus arriving at the Friedmann equation
\be
\frac 1 {a^2} \left(\frac{\text{d}a}{\text{d} \tilde{t}}\right)^2= - \frac \lambda 2 +\frac {2k} {a^4}\; .
\ee
This was the basis of  discussion of  ``BFSS cosmology'' in  \cite{Brahma:2022dsd}.

Although  the change of variables \eqref{dtt} is similar in spirit with our \eqref{dt=2}, for our model with non-constant ${\cal M}$ the complete equation for the change of time variable is much more complicated
\be\label{dtt=}
\text{d}\tilde{t}=  \text{d}\tau  \,  \frac{2a(\tau) }{\mathcal{M}}\frac {\left(1 - \frac{1}{\mu^6}\frac{\mathcal{M'}}{\mathcal{M}}\mathcal{H} \right)} { \left(1 + \frac{1}{\mu^6} \frac{\mathcal{M'}}{\mathcal{M}}\left(2{\cal H} -\frac {3\lambda} {2} a^4(\tau)\right)  \right)}\;   \qquad
\ee
 so that to find the evolution in proper time is not an easy problem.

\end{widetext}

\end{document}